\documentclass[10pt]{ewsn-proc}

\usepackage{booktabs} 
\usepackage[acronym, nowarn]{glossaries}
\makeglossaries

\usepackage{graphicx}
\usepackage{balance}
\usepackage{comment}
\usepackage[font=footnotesize]{caption}
\usepackage{subcaption}
\usepackage{xspace}

\usepackage[hidelinks]{hyperref}

\usepackage{enumitem}

\newcommand{\name}{\textsc{\textrm{HardZiPA}}\xspace}

\usepackage{newtxmath}

\usepackage{makecell}

\usepackage{multirow}

\usepackage{nicematrix}

\definecolor{mygreen}{HTML}{15b01a}
\definecolor{myred}{HTML}{e50000}

\usepackage{tikz}
\newcommand{\tikzcircle}[2][red,fill=red]{\tikz[baseline=-0.5ex]\draw[#1,radius=#2] (0,0) circle ;}%

\newacronym{iot}{IoT}{Internet of Things}
\newacronym{zip}{ZIP}{zero-interaction pairing}
\newacronym{zia}{ZIA}{zero-interaction authentication}
\newacronym{co2}{CO2}{carbon dioxide}
\newacronym{tvoc}{TVOC}{total volatile organic compounds}
\newacronym{prng}{PRNG}{pseudorandom number generator}
\newacronym{hvac}{HVAC}{heating, ventilation, and air conditioning}
\newacronym{rnn}{RNN}{recurrent neural network}
\newacronym{ltsm}{LTSM}{long short-term memory}
\newacronym{wpm}{wpm}{words per minute}
\newacronym{dtw}{DTW}{dynamic time warping}
\newacronym{tfd}{TFD}{time-frequency distance}
\newacronym{dos}{DoS}{denial-of-service}
\newacronym{far}{FAR}{False Acceptance Rate}
\newacronym{frr}{FRR}{False Rejection Rate}
\newacronym{eer}{EER}{Equal Error Rate}

\numberofauthors{3}
\author{
\alignauthor Mikhail Fomichev \\
  \affaddr{Technical University of Darmstadt} \\
  \email{mfomichev@seemoo.tu-darmstadt.de} \\
\alignauthor Timm Lippert \\
  \affaddr{Technical University of Darmstadt} \\
  \email{timm.lippert@gmail.com}
\alignauthor Matthias Hollick \\
  \affaddr{Technical University of Darmstadt} \\
  \email{mhollick@seemoo.tu-darmstadt.de} 
}

\title{Hardening and Speeding Up Zero-interaction Pairing and Authentication}

\begin{document}

\maketitle

\begin{abstract}

Establishing and maintaining secure communications in the \gls{iot} is vital to protect smart devices.
\textit{\Gls{zip}} and \textit{\gls{zia}} enable \gls{iot} devices to establish and maintain secure communications without user interaction by utilizing devices' \textit{ambient context}, e.g., audio.
For autonomous operation, \gls{zip} and \gls{zia} require the context to have enough entropy to resist attacks and complete in a timely manner.
Despite the low-entropy context being the norm, like inside an unoccupied room, the research community has yet to come up with \gls{zip} and \gls{zia} schemes operating under such conditions. 
We propose \name, a novel approach that turns commodity \gls{iot} actuators into injecting devices, generating high-entropy context.
Here, we combine the capability of \gls{iot} actuators to impact the environment, e.g., emitting a sound, with a \gls{prng} featured by many actuators to craft hard-to-predict context stimuli.
To demonstrate the feasibility of \name, we implement it on off-the-shelf \gls{iot} actuators, i.e., \textit{smart speakers}, \textit{lights}, and \textit{humidifiers}. 
We comprehensively evaluate \name, collecting over \textit{80 hours} of various context data in real-world scenarios. 
Our results show that \name is able to thwart \textit{advanced active attacks} on \gls{zip} and \gls{zia} schemes, while doubling the amount of context entropy in many cases, which allows \textit{two times faster} pairing and authentication.

\end{abstract}

%
%

%



\glsresetall

\section{Introduction}
\label{sec:intro}
The proliferation of the \gls{iot} urges the need to \textit{establish and maintain} secure communications between smart devices to protect not only the data, which they exchange, like sensor readings, but also devices themselves, e.g., from unauthorized access.
This ensures the user \textit{privacy} and \textit{trustworthiness} of \gls{iot} systems~\cite{Fomichev:2021, Han:2018, Karapanos:2015, Truong:2014}.
Device \textit{pairing} and \textit{authentication} are approaches to (1) establish a secure communication channel between two devices, and (2) maintain it afterwards. 
By means of pairing two unassociated devices derive a \textit{shared secret key} without any trusted third party, bootstrapping their secure channel~\cite{Fomichev:2017, xu2021key}, while authentication allows one device to assure the legitimacy of another device on such a channel~\cite{conti2020context, mayrhofer2020adversary}.   

Traditionally, pairing and authentication rely on \textit{user interaction} (e.g., entering a password) to fulfill their purposes. 
Yet, the rapidly increasing number of \gls{iot} devices demands a \textit{prohibitive user effort} to pair and authenticate them \cite{Karapanos:2015, Schurmann:2017}.
Even worse, many \gls{iot} devices lack user interfaces, making user-assisted pairing and authentication infeasible~\cite{conti2020context, Fomichev:2021}. 
To address these issues, research proposes \textit{\gls{zip}} and \textit{\gls{zia}}~\cite{Karapanos:2015, Miettinen:2014}.
They allow \textit{colocated devices}, that reside inside an enclosed physical space, e.g., a room, to pair and authenticate \textit{without user involvement}, utilizing devices' ambient context, like audio, captured by their on-board sensors~\cite{Schurmann:2017, Truong:2014}. 
Despite \gls{zip} and \gls{zia} offer  improved \textit{usability} by minimizing user interaction and \textit{deployability} by using off-the-shelf sensors of \gls{iot} devices, their \textit{security} relies upon unpredictability of context, which depends on the intensity and variety of ambient activity (e.g., sound) happening in the environment~\cite{Fomichev:2021}. 

Recent studies scrutinizing \gls{zip} and \gls{zia} schemes find that 
the \textit{insufficient entropy of context}, which is common in many scenarios, like an unoccupied smart home, results in \textit{attacks} against the schemes while prolonging their \textit{time to complete} pairing and authentication~\cite{Fomichev:2019perils, Bruesch:2019,  Shrestha:2018, shrestha2016sounds}.
To date, there exist \textit{few solutions} to address such issues, stemming from the low entropy of context, in the domain of \gls{zip} and \gls{zia}.
We review these solutions in~\autoref{sec:bkgrd}.  

In this work, we propose \name---a novel approach that allows \gls{zip} and \gls{zia} to prevent advanced active attacks while shortening schemes' completion time. 
The idea behind \name is simple, yet powerful: we exploit off-the-shelf \gls{iot} actuators, like a robotic vacuum cleaner, to \textit{generate context which is hard to predict}.
We are inspired by the ubiquity of \gls{iot} actuators found in home and office spaces, e.g., voice assistants, smart lights, and cleaning robots~\cite{maiti2019light, Han:2018}.
Unlike prior \gls{zip} and \gls{zia} works that relied on a human to generate context, like by walking or talking~\cite{Schurmann:2017, Karapanos:2015}, we are the first, to the best of our knowledge, to research the applicability of \textit{commodity \gls{iot} actuators} to produce hard-to-predict context. 
Our motivation for utilizing \gls{iot} actuators is twofold: (1) in many scenarios, e.g., a smart home or office, \textit{humans can be absent} for extended amounts of time, like from home during working hours or from office during lunch breaks, rendering the scenario context predictable due to low entropy~\cite{Shrestha:2018, Han:2018};
(2)  human actions that affect context, e.g., motion or speech, are sufficiently \textit{deterministic}, allowing an adversary to either approximate the context of legitimate devices~\cite{Bruesch:2019, Shrestha:2018, shrestha2016sounds} or even dominate it, like by playing a loud sound~\cite{Han:2018, Fomichev:2019perils}.

Considering our motivation, we focus on \textit{unattended scenarios},\footnote{As part of our system model (cf.~\autoref{sec:mod}), we exemplify why unattended scenarios are relevant for \gls{zip} and \gls{zia} schemes.} e.g., a smart home without any residents, to evaluate \name for two reasons: (1) \textit{attacks} against \gls{zip} and \gls{zia} schemes are \textit{feasible} in these scenarios due to adversary's reduced effort to guess and manipulate context, as well as the lack of legitimate users who can notice the attack~\cite{Shrestha:2018, shrestha2016sounds, conti2020context};
(2) human absence \textit{lowers the amount of entropy} in context, requiring \gls{zip} and \gls{zia} schemes to accumulate more context data to maintain their security, leading to pairing and authentication time on the order of \textit{minutes or even hours}, which is prohibitive for many \gls{iot} applications~\cite{west2021moonshine, Fomichev:2021}. 
Despite addressing the unattended scenarios, we also study the impact of human presence on the efficacy of \name, providing the first insights into how users may perceive it. 

In \name, we leverage (1) the capability of \gls{iot} actuators to ``stimulate'' the context as well as (2) the presence of a \gls{prng} on many actuators~\cite{kietzmann2021guideline}.
Combining these two points, we devise an approach enabling \gls{iot} actuators to \textit{inject hard-to-predict context stimuli}, like audio, into the environment of colocated devices (e.g., a room), boosting the amount of entropy in context, which hardens \gls{zip} and \gls{zia} schemes against attacks and reduces their completion time.  
For realizing \name, we investigate how different context stimuli, like light or audio, can be crafted in a \textit{generic fashion} and then be instantiated on real devices. 
To demonstrate the efficacy of \name, we design and implement it on off-the-shelf \gls{iot} actuators, namely \textit{smart speakers}, \textit{lights}, and \textit{humidifiers}. 
We evaluate \name in real-world home and office scenarios, collecting over \textit{80 hours} of various sensor data, capturing context. 
Our findings reveal that \name thwarts \textit{active attacks} on \gls{zip} and \gls{zia},
which can hardly be prevented by the state-of-the-art schemes~\cite{conti2020context, xu2021key}, while it also increases the amount of context entropy by up to \textit{two times}, allowing \gls{zip} and \gls{zia} schemes to speed up their completion by the same factor. 

In summary, we make the following contributions:
\begin{itemize}
  \item We design \name, a novel approach that leverages \gls{iot} actuators to improve security and shorten the completion time of \gls{zip} and \gls{zia} schemes. 
  \item We implement \name on off-the-shelf \gls{iot} actuators and evaluate it by collecting real-world sensor data, demonstrating the effectiveness of \name. 
  \item We publicly release our collected sensor dataset and the source code of \name implementation. 
\end{itemize}

\section{Background and Related Work}
\label{sec:bkgrd}
We first explain the working principles of \gls{zip} and \gls{zia} schemes and then review related work.

\smallskip
\noindent
\textbf{Background.} 
\gls{zip} and \gls{zia} schemes utilize the \textit{similarity of context} observed by colocated devices (e.g., inside the same room) to either establish a shared secret key or verify devices' physical proximity~\cite{Miettinen:2018, Fomichev:2019perils, Truong:2014}. 
In \gls{zip}, two devices agreeing to pair (1) sense their shared context, like audio, using on-board sensors, (2) translate the captured context into bit sequences called \textit{fingerprints}, and (3) input these fingerprints into a key agreement protocol to establish the shared secret key. 
In \gls{zia}, (1) one device requests an authentication from another device, (2) both devices sense their shared context for a predefined timeframe (e.g., 5 seconds), and (3) the requesting device sends its context readings to the authenticator device, which compares the received context readings with its own, to make the authentication decision.
Note that the two devices performing \gls{zia} are assumed to share a secret key, protecting the transmitted sensor readings from eavesdropping and tampering with---this key can be preloaded or established via device pairing~\cite{conti2020context, Karapanos:2015, Fomichev:2017}.

\smallskip
\noindent
\textbf{Related Work.} 
To date, a few dozen \gls{zip} and \gls{zia} schemes utilizing different sensor modalities, like audio and illuminance, to capture context have been proposed---these are surveyed in~\cite{xu2021key} and~\cite{conti2020context}.  
Several studies disclose the perils of low-entropy context, reducing the security of \gls{zip} and \gls{zia} schemes while prolonging their completion time~\cite{Fomichev:2019perils, Bruesch:2019, Shrestha:2018, shrestha2016sounds}. 
Currently, few solutions exist to address these issues.
\textit{FastZIP} presents a novel \gls{zip} architecture resisting advanced attacks (e.g., similar-context attack) and shortening the pairing time simultaneously~\cite{Fomichev:2021}.
However, this scheme needs various sensors to capture context, decreasing its deployability. 
\textit{Moonshine} distills \gls{zip} fingerprints obtained from low-entropy context, to output the high-entropy fingerprints~\cite{west2021moonshine}. 
While this approach improves the security, it operates by discarding parts of the fingerprint that have low entropy.
Hence, the fingerprints input to Moonshine need to contain redundancy (i.e., more bits), which is achieved by collecting extra context data, thus increasing the pairing time. 

In \gls{zia}, \textit{DoubleEcho} utilizes sound emission, allowing colocated devices to observe a similar room impulse response, which captures unique characteristics of the environment~\cite{Truong:2019}.
While being conceptually close to \name, DoubleEcho focuses on preventing strong \textit{colocated adversaries}, whom we do not consider (cf.~Section~\ref{sec:mod}).
Such stringent security comes at the expense of the scheme's practicality, i.e., DoubleEcho only works for colocated devices residing within half a meter distance. 
\textit{Proximity-Proof} is similar to DoubleEcho regarding the used audio context and threat model, but demands colocated devices to have both a speaker and microphone, reducing the scheme's applicability~\cite{han2018proximity}.

We view \name as a complementary contribution to the above \gls{zip} and \gls{zia} schemes.
Despite sharing some of their weaknesses, like the need for \gls{iot} actuators similar to extra devices and sensors, \name improves the security and pairing time of existing schemes (cf. \autoref{sub:eval-cmp}). 
Using ubiquitous \gls{iot} actuators allows generating \textit{various types of context} (e.g., audio, illuminance) accommodating \gls{zip} and \gls{zia} schemes that rely on single-sensor~\cite{Karapanos:2015},  multi-sensor~\cite{Truong:2014, Fomichev:2021}, and heterogeneous~\cite{Han:2018} contexts.  

We find one system---\textit{Listen!}~\cite{mei2019listen}---that has a similar goal to \name. 
While this research is an important preliminary work for context stimuli injection, there are two notable differences with \name: 
(1) Listen! is customized for audio context injection, whereas \name is a generic approach extensible to diverse context stimuli (cf. \autoref{sec:sysdes}); (2) \name is evaluated under a stronger threat model, where an active adversary can inject their stimuli via an open door (cf.~\autoref{sec:mod}), while in Listen! such an attacker is separated from legitimate devices by a solid wall.


\section{System and Threat Models}
\label{sec:mod}
We present our system model, detailing the goal, requirements, and assumptions of \name, as well as our threat model, describing the goal and capabilities of the adversary.

\smallskip
\noindent
\textbf{System Model.} 
The \textit{goal} of \name is to inject hard-to-predict context stimuli, such as audio, into the environment (e.g., a room), where colocated devices perform \gls{zip} or \gls{zia}, leveraging off-the-shelf \gls{iot} actuators. 
These stimuli seek to increase the similarity and entropy of context observed by colocated devices, allowing \gls{zip} and \gls{zia} schemes to withstand attacks and speed up their completion time. 
We design \name to fulfill the following \textit{requirements}: (1) perform without user interaction after the start-up (\textit{usability}); (2) execute on off-the-shelf \gls{iot} actuators without adding major  modifications to their software stacks (\textit{deployability}). 
To achieve \name's goal while satisfying its requirements, we make the following \textit{assumptions}: (1) \gls{iot} actuators are trusted, i.e., \textit{not} compromised with malware; (2) they feature cryptographically secure \glspl{prng},\footnote{Most operating systems and programming languages have such \glspl{prng} built-in, e.g., \textit{random} in Linux and \textit{secrets} in Python.} producing \textit{unpredictable} random numbers suitable for security purposes~\cite{kietzmann2021guideline}. 

\textit{Unattended Scenarios.} Despite \gls{zip} and \gls{zia} are initiated by a user, there exist cases when devices decide to (re-)pair or (re-)authenticate themselves. 
For example, if there are many devices, \gls{zip} may need a few pairing iterations to ensure confidence in the shared key to exclude passing by devices~\cite{Miettinen:2014};
meanwhile, a user who initiated such \gls{zip} can leave, trusting that pairing will succeed. 
Another case is when devices repair or reauthenticate if one of them gets compromised or as part of a key refreshing routine~\cite{west2021moonshine}. 
Finally, \gls{iot} robots can autonomously change their location (e.g., different rooms inside a smart building), prompting the need to pair or authenticate with in-room devices, irrespective of user's presence.  

We envision \name to be mainly used in \textit{unattended scenarios}, like a smart home without occupants but \gls{iot} actuators within, which challenge security and completion time of \gls{zip} and \gls{zia} schemes~\cite{mei2019listen, Fomichev:2019perils, Shrestha:2018, Miettinen:2018}. 
To produce hard-to-predict context, \name can be invoked periodically or on-demand, based on the needs of \gls{zip} and \gls{zia} schemes. 

\smallskip
\noindent
\textbf{Threat Model.} 
The \textit{goal} of an adversary is to break a \gls{zip} or \gls{zia} scheme by pairing or authenticating with legitimate (i.e., colocated) devices while residing \textit{outside} their environment, such as a room.
To accomplish this goal, the adversary tries to obtain context readings akin to that of legitimate devices', utilizing similar sensing hardware. 
The adversary carries out either a \textit{passive} or \textit{active attack} to achieve their goal.   
In the former attack, the adversary is located right outside the environment of legitimate devices, like in front of a \textit{closed door} to a room; this threat model is followed by most of \gls{zip} and \gls{zia} schemes~\cite{Truong:2014, Miettinen:2018, Han:2018, lee2019voltkey, Karapanos:2015}. 
In the latter attack, the adversary, on top, uses a \textit{half-open door} to the environment of legitimate devices to actively inject their own context stimuli (e.g., audio), employing commodity \gls{iot} actuators or household appliances.
We note that active attacks are difficult to prevent by existing \gls{zip} and \gls{zia} schemes~\cite{Bruesch:2019, Shrestha:2018, mei2019listen}. 

We consider \textit{colocated adversaries} and \textit{\gls{dos} attacks} to be outside the scope of this work. 
While the former is the most difficult attack to defend against in \gls{zip} and \gls{zia}, colocated adversaries (e.g., in the same room as legitimate devices), by definition, undermine the notion of physical security assumed in many use cases, like a smart home~\cite{Han:2018}. 
Despite \gls{dos} attacks being feasible for \gls{zip} and \gls{zia}~\cite{Fomichev:2019perils}, they can only prevent pairing and authentication but \textit{not} circumvent them, leading to a false sense of security. 


\section{System Design and Implementation}
\label{sec:sysdes}
We detail the design and implementation of \name. 

\begin{figure}
\centering
  \includegraphics[width=0.7\linewidth]{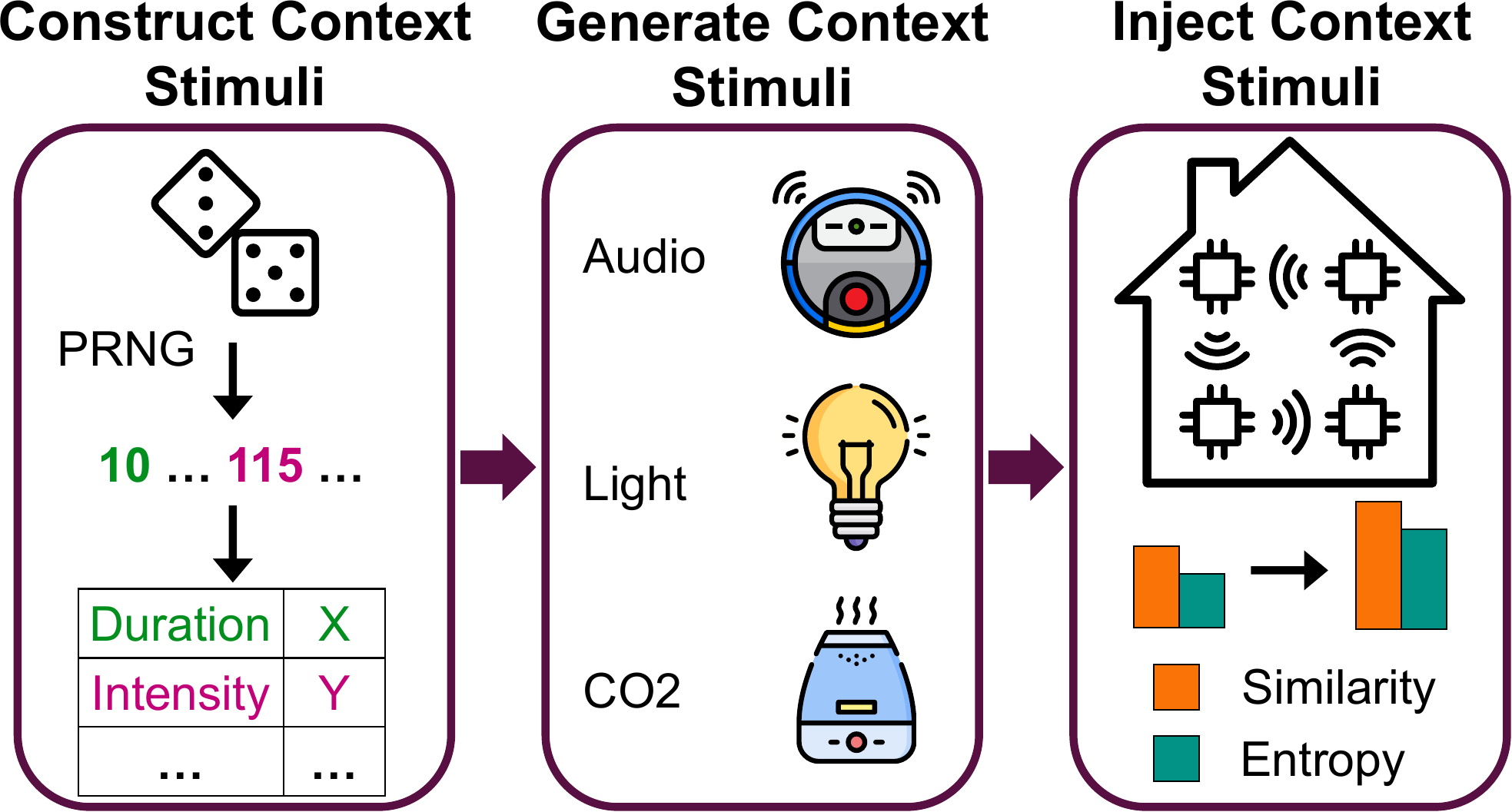}
  \caption{\name injects hard-to-predict context stimuli into the environment of colocated devices (e.g., inside a room) performing \gls{zip} or \gls{zia} to increase the similarity and entropy of context observed by these devices. \name constructs such stimuli by shaping their form and occurrence using a \gls{prng} and generates the stimuli utilizing off-the-shelf \gls{iot} actuators.}
  \label{fig:sys-design}
  \vspace{-3.5ex}
\end{figure}

\smallskip
\noindent
\textbf{System Overview.} 
\name produces hard-to-predict context stimuli to \textit{secure} \gls{zip} and \gls{zia} schemes against attacks and \textit{shorten} their completion time as follows (cf.~\autoref{fig:sys-design}). 
First, \name constructs a context stimulus that can be produced by a specific actuator, such as audio by the speaker.
For this, we use a \gls{prng} to randomize the stimulus parameters, e.g., frequency, duration, intensity, and pattern (cf.~\autoref{tab:stim-params}). 
Second, different off-the-shelf \gls{iot} actuators generate context stimuli constructed in this fashion, affecting various types of context, like light or audio. 
Third, when launched in the environment, where colocated devices execute a \gls{zip} or \gls{zia} scheme (e.g., a room), \name continuously injects context stimuli for the duration required by the scheme (cf.~\autoref{sec:mod}), increasing the similarity and entropy of context captured by these colocated devices. 

\subsection{Context Stimuli Injection Algorithm}
\label{sub:stim-alg}
Before describing how \name injects stimuli into the context, we first rationalize the used \gls{iot} actuators influencing various types of context captured by specific sensors.   

\smallskip
\noindent
\textbf{Selected  \gls{iot} Actuators.}
We design \name in a generic way, choosing various \gls{iot} actuators that (1) are ubiquitous and (2) can impact specific context (e.g., audio) utilized by existing \gls{zip} and \gls{zia} schemes. 
Based on these two criteria, we select a \textit{smart speaker}, \textit{light}, and \textit{humidifier} as actuators to implement and evaluate \name. 
The smart speakers are on the rise due to the proliferation of voice assistants, like Alexa, while being built into many other smart devices, e.g., vacuum cleaning robots.  
Hence, the \textit{audio} played by speakers can be captured by microphones of multiple user-end \gls{iot} devices, like smartphones or watches. 
It is thus unsurprising that the audio context is most frequently used by existing \gls{zip} and \gls{zia} schemes~\cite{Miettinen:2018, Karapanos:2015, Truong:2014, Truong:2019, han2018proximity, Han:2018}.

The popularity of smart lights is increasing due to energy-saving and sustainability concerns~\cite{maiti2019light}. 
The \textit{illuminance} produced by smart bulbs can be captured by low-power ambient light sensors that are pervasive in \gls{iot} devices, while the \textit{light's color} can be recorded by RGB sensors, which are also widespread~\cite{xu2021key}.
Hence, the light context has been extensively utilized by \gls{zip} and \gls{zia} schemes~\cite{Miettinen:2014, liu2017secure, lu2020telling}. 

\begin{table}
\scriptsize
	\centering
	\caption{Parameters of context stimuli used in \name.}
	\label{tab:stim-params}
	\begin{tabular}{lccc}
		\toprule
		\multirow{2}[2]{*}{\shortstack[l]{Stimuli \\ parameter}} & \multicolumn{3}{c}{Context stimuli} \\
		\cmidrule(rl){2-4}
		& Audio & Light & CO2 \\
		\midrule
		Frequency & 1--3 min & 30--90 sec & 5--10 min \\
		Duration & 4.5--75 sec & 5--60 sec & 10--15 min \\ 
		Intensity & \makecell{Loudness: \\ 0.1--1} & \makecell{Brightness: \\ 1--254} & \makecell{Mist emission: \\ low/high}\\
		Pattern & Speech + noise & Blink/constant & n/a\\
		Color & n/a & RGB: 0--255 & n/a \\
		Spectrum & 50--22100 Hz & n/a & n/a \\
		\bottomrule
	\end{tabular}
	\vspace{-3.5ex}
\end{table}

The number of smart humidifiers is skyrocketing, boosted by the Covid-19 pandemic, as they can reduce viruses spread and enable a healthier in-door environment~\cite{hum-covid:2020}. 
A humidifier emitting vapor affects not only the \textit{humidity} of the environment but also its \textit{temperature} and \textit{\gls{co2}} concentration~\cite{liu2017study}. 
These three modalities can be captured by integrated environmental sensors that become ubiquitous~\cite{Bosch:2021}.  
We identify one \gls{zia} scheme that relies on humidity, temperature, and \gls{co2} contexts~\cite{Shrestha:2014}. 
However, recent concerns for public health and climate protection should prompt the massive adoption of environmental sensors, enabling future \gls{zip} and \gls{zia} schemes. 
In \name, we focus on the \gls{co2} context due to its major importance for human health as well as pollution and climate monitoring. 

\smallskip
\noindent
\textbf{Overview of Context Stimuli Injection.}
To produce hard-to-predict context stimuli with \name, we rely upon the following two observations: (1) actions performed by actuators, i.e., context stimuli, are \textit{configurable} like sound loudness; and (2) many \gls{iot} actuators \textit{feature a \gls{prng}} for functional and security purposes~\cite{kietzmann2021guideline}. 
Thus, by using the \gls{prng}, we are able to parameterize the actions of \gls{iot} actuators to be non-deterministic, generating hard-to-predict context stimuli. 
However, a \textit{main challenge} here is to identify the stimuli parameters that are generic for various \gls{iot} actuators and set them to yield \textit{higher context similarity} and \textit{entropy} observed by colocated devices in their environment, e.g., a room.  
Table~\ref{tab:stim-params} shows such stimuli parameters and their values used in \name. 
In the following, we explain how these parameters have been selected based on our empirical findings.

The \textit{algorithm} for context stimuli injection in \name works as such (cf.~\autoref{fig:sys-design}): (1) each actuator uses its \gls{prng} to produce the sequence of random numbers, where (2) every random number controls the assigned stimuli parameter, like duration in~\autoref{tab:stim-params}, to shape the context stimulus executable by this actuator (e.g., audio by speakers), which (3) generates such crafted stimulus, repeating the complete algorithm, i.e., steps (1)--(3), after a random pause. 

\smallskip
\noindent
\textbf{Audio Injection.} 
We use smart speakers to inject audio stimulus. 
Since microphones have a short response time, sensing sound almost instantaneously, we set the duration of stimulus to 4.5--75 seconds, allowing for both shorter and longer audio injections. 
The duration parameter \textit{impacts} the frequency of stimulus occurrence: we choose it 1--3 minutes to keep a balance between the time required by the speaker to inject the stimulus and its regularity. 
We express the intensity of audio as loudness, normalized between 0.1--1, reflecting the minimum and maximum \textit{loudness output} by a specific speaker.   
As for the pattern of stimulus, we combine \textit{speech and noise}; the latter is generated in a wide frequency spectrum from 50 Hz to 22 kHz, which can be produced and captured by off-the-shelf speakers and microphones, respectively~\cite{Karapanos:2015}.

\begin{figure}
\centering
  \includegraphics[width=0.85\linewidth]{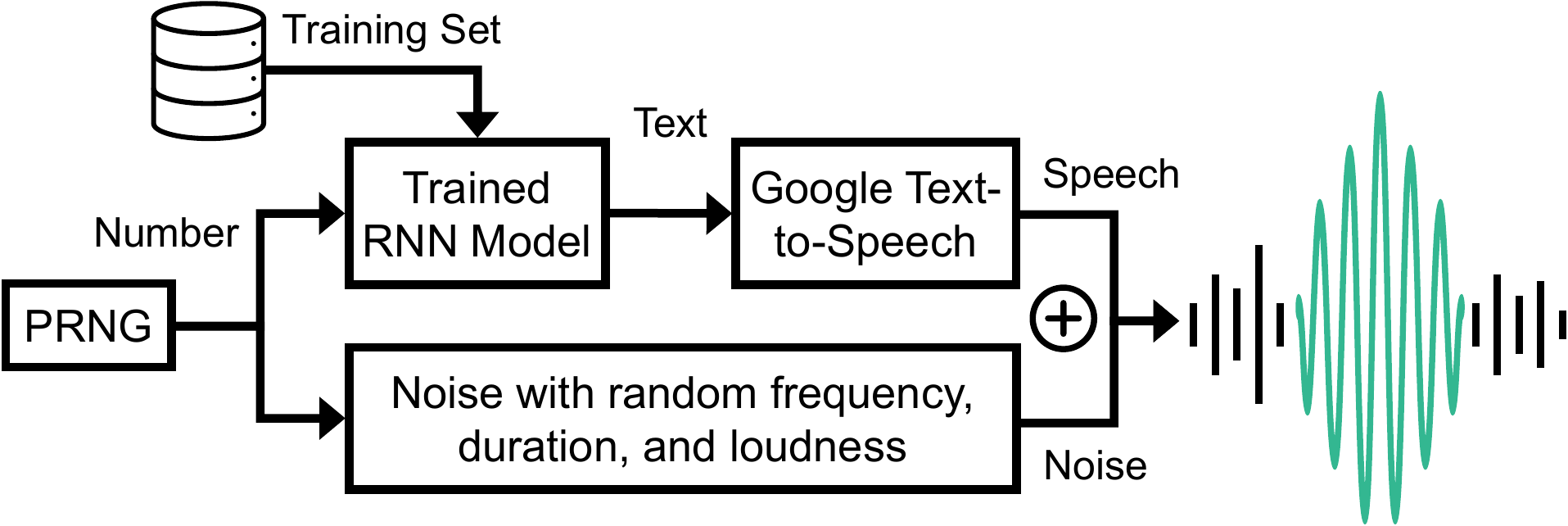}
  \caption{Structure of \name's audio stimulus generator. It produces sound comprised of the randomized speech which is interleaved with noise signals of random frequency, duration, and loudness.}
  \label{fig:audio-gen}
  \vspace{-3.5ex}
\end{figure}

Figure~\ref{fig:audio-gen} shows the structure of \name's audio stimulus generator. 
To produce speech, we use a real-world text corpus\footnote{Contains 96 sentences in English with minimum 2, average 7.12, and maximum 30 words~\cite{dustcloud}.} from the robotic vacuum cleaner, which is one of the actuators in our experiments (cf.~\autoref{sub:exp-setup}).  
Based on this corpus, we train a \gls{rnn} to generate random text; such a technique is popular~\cite{fedus2018maskgan}. 
Our \gls{rnn} has \textit{three layers} as inspired by~\cite{OurRNN:2021}: (1) embedding with 256 dimensions, (2) \gls{ltsm} with 1024 units, and (3) dense.
We set the \gls{rnn} \textit{temperature parameter}  to 0.45, yielding the best trade-off between unpredictability and comprehensibility of a text generated by our model, that we implement in Keras. 
Then, such generated text is input to the Google's Text-to-Speech converter~\cite{T2S:2021},  producing speech at the conversational rate of 120 \gls{wpm}~\cite{Speech:2018}. 

To further increase the \textit{unpredictability} of audio injection, we insert into the speech 1--3 evenly distributed noise signals with varying duration (0.5--5 seconds), loudness (0.1--1), and spectrum (50--22100 Hz).
The number of inserted noise signals depends on the speech length, which is found as the duration of one word, i.e., 2 seconds at 120 \gls{wpm}, multiplied by the word count in the \gls{rnn} generated text, i.e., 2--30 words. 
Other noise parameters---duration, loudness, spectrum---are chosen randomly using the \gls{prng} output.
To select the shape of noise signal, we experiment with sine, square, and sawtooth waveforms, finding that the latter two introduce harmonics~\cite{Waveforms:2018}, which complicates the comparison of audio contexts captured by devices (cf.~\autoref{sub:sim-ent}). 
Thus, we choose the \textit{sine waveform}, allowing us to generate noise at a specific frequency, such as 500 Hz, without any harmonics.

\begin{figure}
	\centering
	\begin{subfigure}[b]{.45\linewidth}
		\centering
		\includegraphics[width=\linewidth]{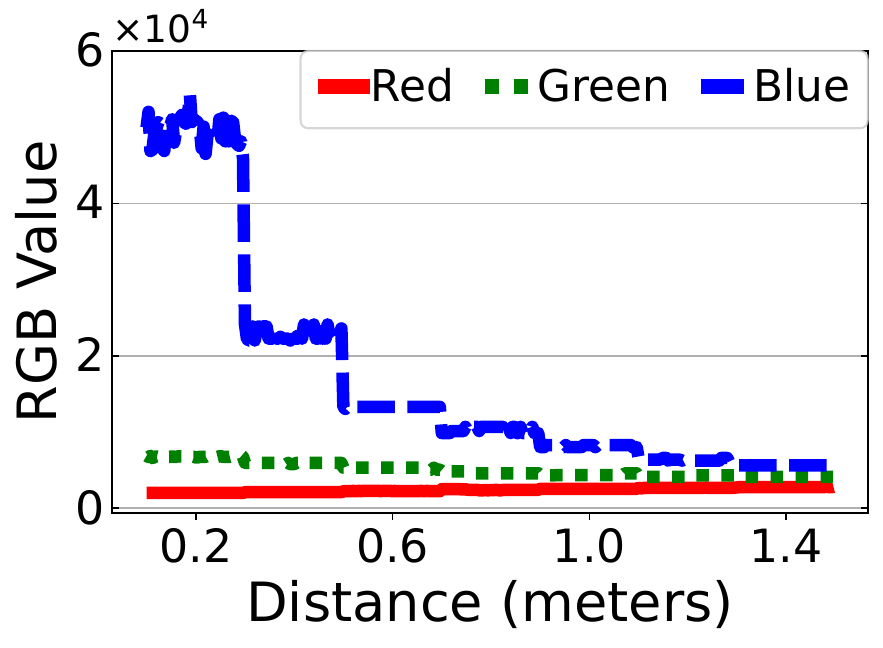} 
		\caption{RGB sensor: smart bulb}
		\label{sf:light_stimuli}
	\end{subfigure}
	\begin{subfigure}[b]{.45\linewidth}
		\centering
		\includegraphics[width=\linewidth]{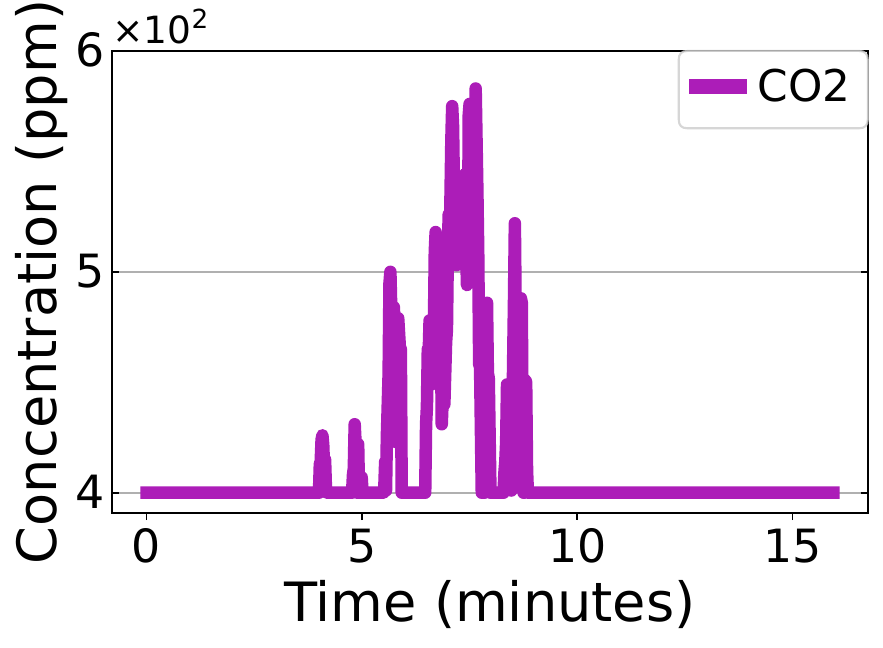}
		\caption{CO2 sensor: humidifier}
		\label{sf:gas_stimuli}
	\end{subfigure}
	\caption{Impact of (a) colored light produced by a smart bulb and (b) water vapor produced by a humidifier on RGB and \gls{co2} sensors, respectively.}
	\vspace{-3.5ex}
\end{figure}

\smallskip
\noindent
\textbf{Light Injection.} 
We utilize smart bulbs to inject light stimulus in terms of illuminance and light color.  
As both ambient light and RGB sensors respond within several milliseconds, we set the stimulus duration to 5--60 seconds to enable brief and extended light injections. 
Given this duration, we inject the stimulus with the frequency of 30 to 90 seconds, ensuring that our hardware can perform it correctly while running uninterrupted. 
We represent the intensity of light as \textit{brightness}, which often has actuator-specific ranges (e.g., 1--254)---these can, in principle, be unified, like from 0 to 1. 
For the pattern of stimulus, we consider \textit{constant and blinking light}, finding that commodity smart bulbs cannot blink faster than once per second.
Hence, we choose the blinking frequency from the interval 0.2--1 Hz, 
providing both reliable light injection and various blinking modes.

To set the color of light, we randomly select R, G, and B values from their 0--255 range, mapping the resulting color onto the \textit{color space of a smart bulb}, i.e., CIE 1931 space in our experiments (cf.~\autoref{sub:exp-setup}). 
\autoref{sf:light_stimuli} shows how colored light generated by \name impacts the RGB sensor that we use. 
This sensor returns undocumented RGB values when being exposed to colored light. 
Therefore, we reverse-engineered its working principle, allowing us to reliably distinguish between red, green, and blue colors in the range of distances from the source of light to the sensor. 

\smallskip
\noindent
\textbf{\gls{co2} Injection.} 
We employ a smart humidifier to inject \gls{co2} stimulus, i.e., by evaporating water, the humidifier changes the \gls{co2} concentration in the environment.
\autoref{sf:gas_stimuli} depicts this process using the off-the-shelf humidifier and \gls{co2} sensors (cf.~\autoref{sub:exp-setup}). 
Here, the humidifier starts to operate after minute 1, working for about five minutes. 
We see that (1) it takes 2--3 minutes for the stimulus to affect the sensor; (2) the \gls{co2} concentration gradually returns to its ambient level once the stimulus is removed.
Based on these findings, we set the stimulus frequency and duration to 5--10 minutes and 10--15 minutes, respectively.
This allows the \gls{co2} stimulus to reach multiple sensors in the environment and settle back to the background level of \gls{co2}.
The humidifier used in our experiments has two levels of intensity: low or high \textit{mist emission}.
We omit setting the pattern of stimulus, as our humidifier features a single nozzle that allows spraying mist in \textit{only one direction}. 
Yet, this stimulus pattern can be realized on advanced humidifiers with multiple mist intensity levels and nozzles, making the \gls{co2} context more unpredictable. 

\subsection{Context Similarity and Entropy}
\label{sub:sim-ent}
We first justify the similarity metrics to compare context captured by two devices and then present our method for estimating the amount of entropy in context. 

\smallskip
\noindent
\textbf{Similarity Metrics.}
In \name, we want to compare the similarity of different types of context, like audio or \gls{co2}, in a \textit{generic way} to abstract from specifics of concrete \gls{zip} and \gls{zia} schemes, whose context similarity metrics often depend on the use case (e.g., smart home vs. wearables) and implementation~\cite{Fomichev:2019perils, west2021moonshine}. 
To achieve this, we consider the \textit{\gls{dtw}} algorithm which measures the distance between two time series, like sensor readings, handling misaligned and different-size data~\cite{lee2020ivpair}. 
We discover that \gls{dtw} suits well for comparing the similarity of illuminance, RGB, and \gls{co2} data---confirming the results of prior work~\cite{maiti2019light}; in \name, we use \gls{dtw}'s Python implementation~\cite{giorgino2009computingDTW}. 

We find \gls{dtw} to be less practical to compare audio data, as it imposes high computational overhead while only allowing the comparison in the time domain. 
Thus, we experiment with other metrics measuring \textit{audio similarity} proposed by prior research: similarity score, maximum cross-correlation, and time-frequency distance~\cite{Karapanos:2015, Truong:2014}. 
Our findings show that the first metric can best capture audio similarity in both time and frequency domains simultaneously, and it behaves stably across different environments (e.g., rooms); these results are in accord with~\cite{Fomichev:2019perils}. 
Hence, we leverage the \textit{similarity score}, which is found as the average of maximum cross-correlations computed in each of 20 one-third octave bands from 50 Hz to 4 kHz, to compare the similarity of two audio recordings.
In \name, we use the Matlab implementation of the similarity score metric from~\cite{Fomichev:2019perils}. 

\smallskip
\noindent
\textbf{Entropy Estimation.}
There exist two ways to estimate entropy in the \gls{zip} and \gls{zia} domain: (1) is to apply NIST tests on the fingerprints, i.e., sequences of bits, generated by \gls{zip} schemes from context data~\cite{turan2018recommendation}; (2) is to compute the amount of entropy in raw context data using its distribution~\cite{west2021moonshine}.
The former method is \textit{not generic}, because it only works for \gls{zip} schemes, whose fingerprints often have \textit{entropy biases} introduced during the translation of context data into the sequence of bits~\cite{Bruesch:2019, Fomichev:2019perils}, while the NIST tests can assess the entropy inaccurately if the amount of input data (e.g., \gls{zip} fingerprints) is insufficient.  
Thus, we adopt the latter approach and utilize the following formula to calculate the amount of \textit{entropy in raw sensor data} that captures context:
\begin{equation}\label{eq:1}
H(X) = -\dfrac{\sum_{i=1}^{b}P(x_{i})\log_{2}P(x_{i})}{\log _{2} b}
\end{equation}
Here, $H(X)$ is the entropy of time series $X$ (i.e., sensor data), treated as a random variable.  
To use this representation, we quantize sensor readings into the $b$ number of bins, obtaining the distribution of such data. 
Therefore, $x_{i}$ is the center of the bin, $P(x_{i})$ is the probability that the random variable $ X $ falls inside bin $i$, while $\log _{2} b$ acts as a normalization factor.

\begin{figure}
	\centering
	\begin{subfigure}[b]{.46\linewidth}
		\centering
		\includegraphics[width=\linewidth]{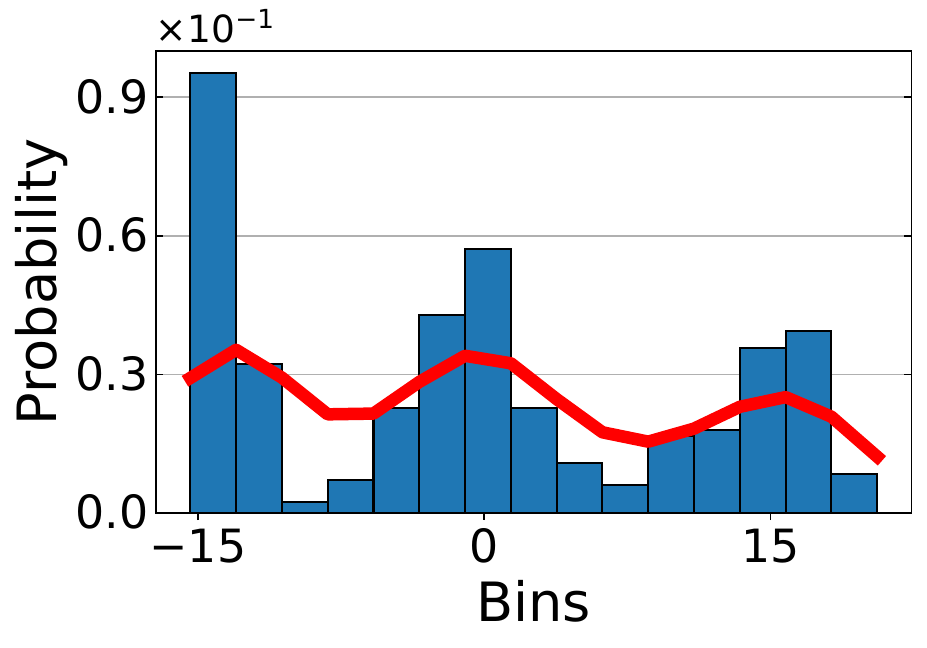}
		\caption{\name works (data $Y$)}
		\label{sf:hist_high}
	\end{subfigure}
	\begin{subfigure}[b]{.46\linewidth}
		\centering
		\includegraphics[width=\linewidth]{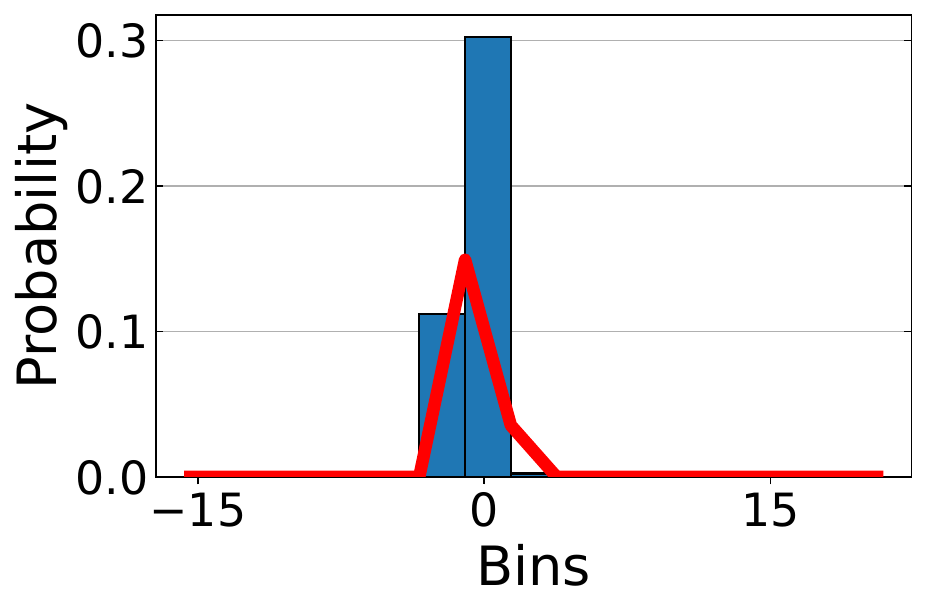} 
		\caption{\name is idle (data $Z$)}
		\label{sf:hist_low}
	\end{subfigure}
		\caption{Distribution of RGB data in two cases, i.e., (a) \name works, and (b) it is idle. In the former case, data $Y$ has a significantly more uniform distribution than in the latter case of data $Z$, hence $Y$ contains more entropy.}
	\label{fig:ent-hist}
	\vspace{-3.5ex}
\end{figure}

\autoref{fig:ent-hist} depicts two distributions obtained in this manner for RGB data when \name (1) works and (2) is idle. 
In the former case (cf.~\autoref{sf:hist_high}), the data distribution is much closer to \textit{uniform} than in the latter case (cf.~\autoref{sf:hist_low}).
Using \autoref{eq:1}, we calculate the entropy of these two time series: $H(Y) = 0.88$ and $H(Z) = 0.23$, which exhibits the impact of \name. 
It is agreed in the \gls{zip} and \gls{zia} domain that the amount of entropy in context data is \textit{directly proportional} to pairing and authentication time~\cite{Karapanos:2015, Fomichev:2021, Han:2018}. 
Intuitively, from more unpredictable context \gls{zip} and \gls{zia} schemes can \textit{faster} extract a distinct fingerprint and context feature, respectively. 
Thus, utilizing data $Y$ allows speeding up \gls{zip} and \gls{zia} by the factor of 3.8, as compared to data $Z$ (cf.~\autoref{fig:ent-hist}). 

To use our entropy estimation method, we need to choose the number of bins for each sensor data, such as \gls{co2}. 
While considering existing bin selection criteria,\footnote{\url{https://www.statisticshowto.com/choose-bin-sizes-statistics/}.} we find that data properties, i.e., range and behavior, to be decisive in choosing the number of bins.  
Based on our empirical findings, we set the following \textit{number of bins} for each sensor data: 19 (audio), 20 (illuminance), 100 (RGB), and 20 (\gls{co2}) to provide consistent and reliable entropy estimation. 

\section{Evaluation of \name}
\label{sec:eval}
We evaluate \name based on the real-world data that we collect to demonstrate its feasibility.

\subsection{Experiment Setup} 
\label{sub:exp-setup}
\noindent
\textbf{Apparatus.}
We prototype \name using the following \gls{iot} actuators. 
To inject audio, we utilize two types of speakers: (1) a built-in speaker of the \textit{Roborock S5} vacuum cleaning robot and (2) a standalone \textit{JBL Flip 5} speaker.
We replicate the procedure in~\cite{dustcloud} to gain access to the Roborock S5, allowing us to play customized audio on it using the \textit{python-miio} package.\footnote{\url{https://github.com/rytilahti/python-miio}.}
The JBL speaker is controlled by connecting it to a laptop that runs the \name logic. 
For light injection, we make use of popular \textit{Philips Hue} lights by connecting smart bulbs with the Hue bridge and managing them via the Philips API~\cite{philips_light_api}. 
To inject \gls{co2}, we leverage an affordable \textit{Maxcio Smart} humidifier, which is commanded through the Tuya API~\cite{TuyAPI}.

Similar to \name, we employ the following \gls{iot} actuators and household appliances to inject adversarial context stimuli during the active attack (cf.~\autoref{sec:mod}): a budget \textit{Lenrue A2} speaker and JBL Flip 5 for audio, a colored lamp and flashlight for light, and a pedestal fan for \gls{co2}.

We build the following platform to collect sensor data. 
To record audio, we utilize a \textit{Samson Go} USB microphone connected to the \textit{Raspberry Pi 3 Model B}.
The audio is captured in a raw PCM16 format at 44.1 kHz sampling rate, and it is stored in a WAV file. 
To collect illuminance and RGB data, we utilize the light sensor of the \textit{Samsung Galaxy S6} smartphone that records these data at 5 Hz sampling rate, streaming it to the Raspberry Pi via USB. 
For \gls{co2} collection, we employ an \textit{SGP30 multigas sensor} attached to the Raspberry Pi via its pins, capturing the data at 1 Hz sampling rate. 
The recordings of our light and gas sensors are stored in text files, where each data point is supplied with a timestamp. 

~\autoref{tab:exp-hardware} summarizes the apparatus, i.e., both actuators and sensors, used in our experiments. 
We note that legitimate and adversarial devices utilize the \textit{same sensing hardware}, which is described above, to collect context data.  

\begin{figure}
\centering
\begin{subfigure}[b]{0.42\linewidth}
	\centering
   	\includegraphics[width=0.98\linewidth]{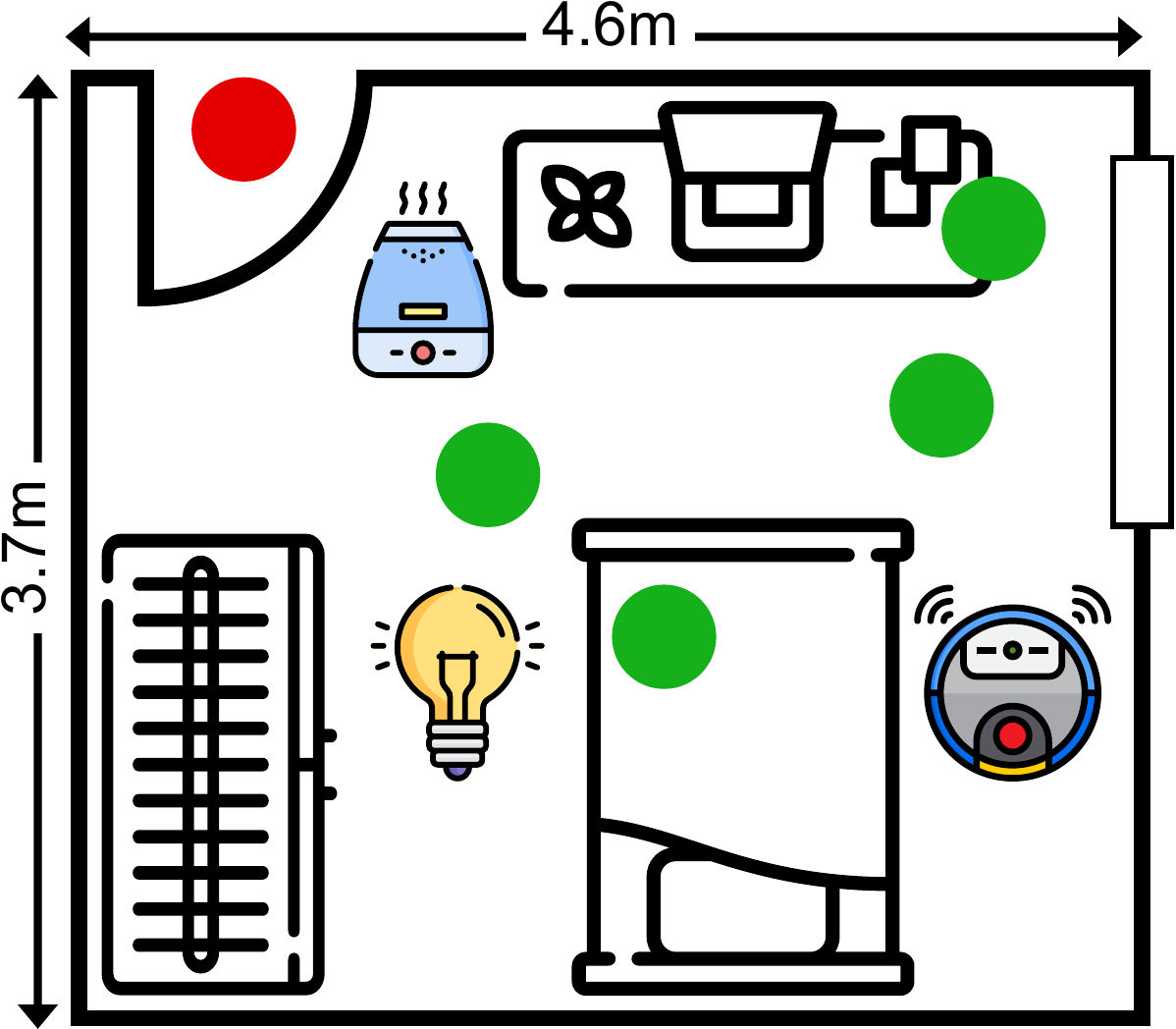}
   	\caption{Home}
   	\label{sf:setup-home}
\end{subfigure}
\begin{subfigure}[b]{0.42\linewidth}
	\centering
   	\includegraphics[width=0.98\linewidth]{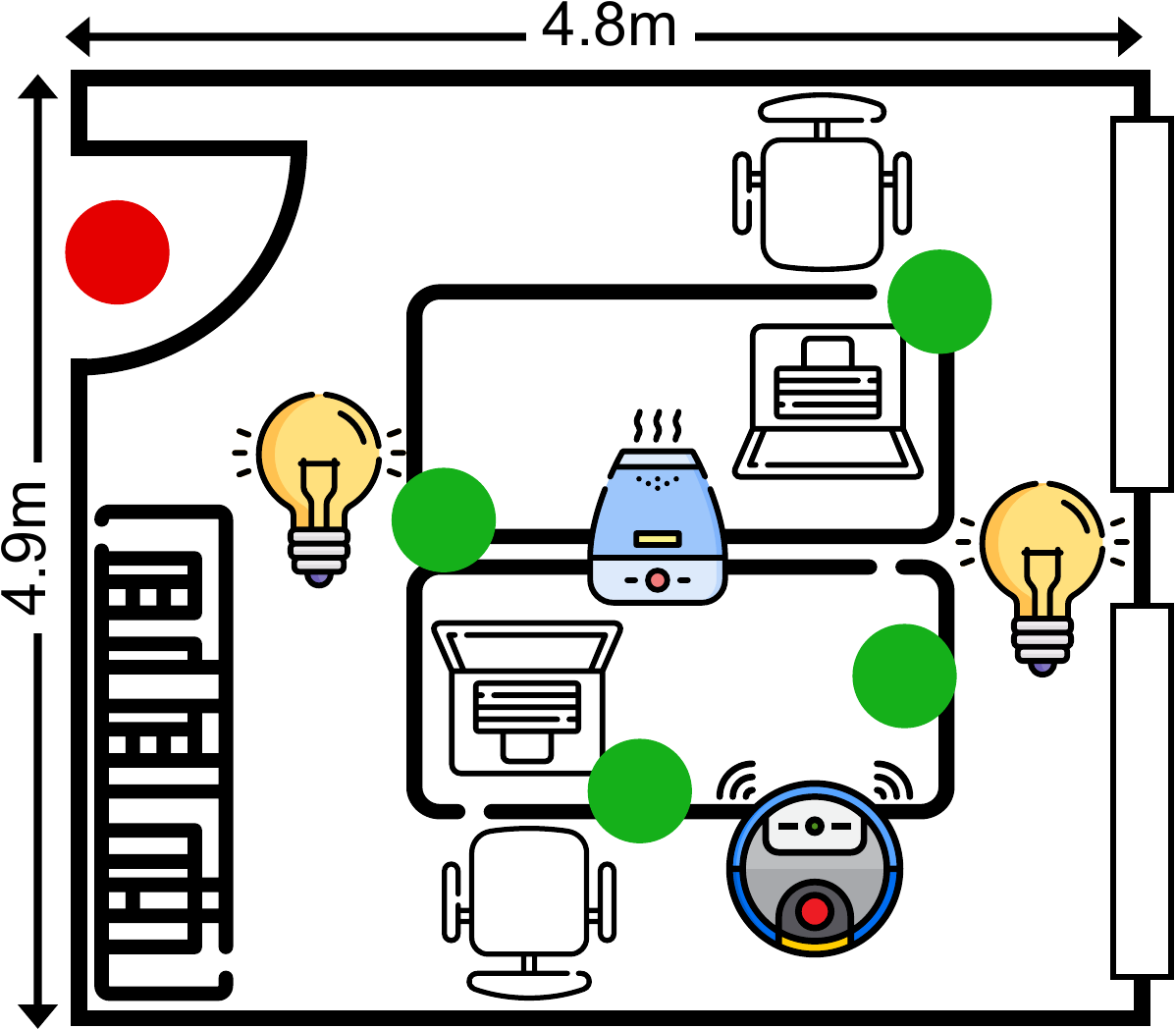}
   	\caption{Office}
    \label{sf:setup-office}
\end{subfigure}
\caption{Data collection settings in our experiments. The legitimate colocated devices (\tikzcircle[mygreen,fill=mygreen]{4.0pt}) and \gls{iot} actuators reside inside a home or office environment, while a non-colocated adversarial device (\tikzcircle[myred,fill=myred]{4.0pt}) stays outside it, i.e., in front of an entrance door which is closed in the passive attack and half-open during the active attack (cf.~\autoref{sec:mod}).}
\label{fig:exp-setup}
\vspace{-3.5ex}
\end{figure}

\smallskip
\noindent
\textbf{Data Collection.}
We capture audio, illuminance, RGB, and \gls{co2} data using our sensing platform in two real-world scenarios: \textit{home} and \textit{office}, as shown in~\autoref{fig:exp-setup}. 
In the former, we deploy four legitimate devices and three actuators (i.e., Roborock S5, two Hue bulbs, Maxcio humidifier) within an apartment room, while the adversarial device resides in front of an entrance door that can be open or closed (cf.~\autoref{sec:mod}). 
In the latter scenario, we equip an office in a similar fashion: with four legitimate devices and one adversarial, but we use four Hue bulbs to account for an office's larger area and more obstacles. 
To further assess the capability of \name to resist attacks and boost context entropy, we experiment with  additional actuators in the office scenario, i.e., the JBL Flip 5 speaker and \textit{iTvanila Mist} humidifier.

\begin{table*}
\scriptsize
  \centering
  	\caption{Apparatus, i.e., actuators and sensors, used for data collection and evaluation of \name.}
	\label{tab:exp-hardware}
	\begin{tabular}{lllc}
		\toprule
		\multirow{2}[2]{*}{\shortstack[l]{Context \\ data}} & 
		\multicolumn{2}{c}{Actuators} & 
		\multirow{2}[2]{*}{\shortstack[c]{Sensors: \textit{same} for legitimate and adversarial devices \\ (sampling data rate)}} \\
		\cmidrule(rl){2-3} 
		& \name & Adversarial & \\
		 \midrule
		Audio & Speaker: Roborock S5 / JBL Flip 5 & Speaker: Lerne A2 / JBL Flip 5 & Microphone: Samson Go USB  (44.1 kHz) \\
		Illum./RGB & Smart light bulbs: Philips Hue & Colored bulbs: lamp / flashlight & Light sensor: Samsung Galaxy S6 (5 Hz) \\
		\gls{co2} & Humidifier: Maxcio Smart / iTvanila Mist & Pedestal fan & Multigas sensor: SGP30 (1 Hz) \\
		 \bottomrule
	\end{tabular}
	\vspace{-3.5ex}
\end{table*}

Following our threat model given in~\autoref{sec:mod}, we collect context data in \textit{four experimental settings}---the first two correspond to passive and active attacks while \name \textit{does not} execute: labeled as \textit{PA} and \textit{AA}, respectively.
The second two settings are for passive and active attacks during which \name \textit{works}, namely \textit{PA+H} and \textit{AA+H}. 
In our evaluation of \name, we focus on \textit{unattended} use cases, i.e., without humans, as motivated by~\autoref{sec:mod}.
Hence, the home scenario remains unoccupied in all experiments (cf.~\autoref{tab:exp-settings}), 
yet we expose it to \textit{realistic ambient conditions}, like external noise from neighbouring rooms / streets, flicker of a display, and drafts from a tilted window. 
While in the office, we introduce two people in the \textit{PA} experiment who actively move and converse.
Thus, we can assess the \textit{impact of human presence} on context and compare it with that of \name. 
The rest of office experiments are unoccupied, but we conduct them under realistic conditions similar to the home (cf.~\autoref{tab:exp-settings}).

To understand whether human presence can interfere with \name, lowering its efficacy, and gain the \textit{first insights} into how users may perceive \name (cf.~\autoref{sec:disc}), we additionally repeat the office \textit{PA+H} and \textit{AA+H} experiments with two people present in the scenario whom we ask to follow their normal working routine, implying occasional conversations and motion of our participants. 
In total, we collect over 80 hours of sensor data in our scenarios. 

\smallskip
\noindent
\textbf{Reproducibility.} Our collected sensor data, intermediate results, and the codebase of \name are publicly available: \textcolor{blue}{\url{https://github.com/seemoo-lab/hardzipa}}. 

\smallskip
\noindent
\textbf{Ethical Considerations.} This research was approved by our institutional review board, the participants residing in experimental settings (cf.~\autoref{tab:exp-settings}) gave informed consent for the collection, use, and release of sensor data.

\begin{table}
\scriptsize
  \centering
  	\caption{Details on \name data collection and evaluation settings.}
	\label{tab:exp-settings}
	\begin{tabular}{lcccc}
		\toprule
		\multirow{2}[2]{*}{\shortstack[l]{Experimental \\ setting}} & \multirow{2}[2]{*}{\shortstack[c]{Attack \\ type}}
		& \multirow{2}[2]{*}{\shortstack[c]{\name \\ works?}} & \multicolumn{2}{c}{Human presence?} \\
		\cmidrule(rl){4-5} 
		& & & Home & Office \\
		 \midrule
		 \textit{PA} & Passive & No & No & Yes \\
		 \textit{AA} & Active & No & No & No \\
		 \textit{PA+H} & Passive & Yes & No & No \\
		 \textit{AA+H} & Active & Yes & No & No \\		 
		 \bottomrule
	\end{tabular}
	\vspace{-3.5ex}
\end{table} 

\subsection{Methodology} 
\label{sub:eval-meth}
\noindent
\textbf{Data Preprocessing.}
To estimate the similarity and entropy of context (cf.~\autoref{sub:sim-ent}), we preprocess our collected sensor data as follows. 
We synchronize the audio recordings of colocated legitimate devices via cross-correlation as in~\cite{Fomichev:2019perils}.  
For illuminance, RGB, and \gls{co2} data, we first perform mean subtraction to eliminate offsets between sensors, e.g., due to hardware variation, and then conduct signal smoothing and noise reduction in two steps: (1) applying a Savitzky-Golay filter using a window length 3 and degree 2 polynomial, followed by (2) a Gaussian filter with a sigma of 1.4.
We adapt the filter parameters for signal smoothing and noise reduction based on best practices from related work~\cite{Fomichev:2021, Lin:2019}.

\smallskip
\noindent
\textbf{Similarity and Entropy Estimation.} 
Recall that we choose the similarity score as our metric to compare audio similarity (cf.~\autoref{sub:sim-ent}).
This metric is best suited for audio snippets \cite{Karapanos:2015}, hence we split our audio data into \textit{recordings} of 10, 30, and 60 seconds.
We then compute the similarity score using such recordings of colocated and non-colocated devices, and find the average result. 
The similarity score ranges \textit{between 0 and 1}, with a bigger number showing higher audio similarity. 
As for illuminance, RGB, and \gls{co2} similarity, we compute it as \gls{dtw} distance between the \textit{full data} of colocated and non-colocated devices collected in each experiment (cf.~\autoref{tab:exp-settings}), finding the average result. 
We normalize our \gls{dtw} distances utilizing min-max scaling to be \textit{from 0 to 1}. 
Here, a smaller distance value indicates higher data similarity.
Moreover, we study \gls{dtw} distances on snippets of illuminance, RGB, and \gls{co2} data to shed light on recording sizes of such modalities that most benefit \gls{zip} and \gls{zia} aided by \name. 

\begin{table}
\scriptsize
  \centering
  	\caption{Similarity and entropy metrics used for \name evaluation.}
	\label{tab:sim-entr-metr}
	\begin{tabular}{llllc}
		\toprule
		\multirow{2}[2]{*}{\shortstack[l]{Context \\ data}} & 
		\multicolumn{3}{c}{Similarity} & 
		\multirow{2}[2]{*}{\shortstack[c]{Entropy: \textit{same} for \\ all context data}} \\
		\cmidrule(rl){2-4} 
		& Metric & Range & High similarity? & \\
		 \midrule
		Audio & Sim. score & $[0, 1]$ & Big sim. score & \multirow{3}{*}{\shortstack[c]{Metric:~\autoref{eq:1} \\ Range: $[0, 1]$}} \\
		Illum./RGB & \gls{dtw} dist. & $[0, 1]$ & Small \gls{dtw} dist. &  \\
		\gls{co2} & \gls{dtw} dist. & $[0, 1]$ & Small \gls{dtw} dist. &  \\
		 \bottomrule
	\end{tabular}
	\vspace{-3.5ex}
\end{table}

To estimate the entropy of audio, illuminance, RGB, and \gls{co2} data, we use~\autoref{eq:1} from~\autoref{sub:sim-ent}. 
Specifically, we input the \textit{full data} of each colocated device (per experiment, cf.~\autoref{tab:exp-settings}) into this equation to calculate its entropy, averaging the results. 
By considering the full data per device, we keep the entropy estimation \textit{generic}, avoiding parameters of specific \gls{zip} and \gls{zia} schemes, e.g., size of the input data. 
Our resulting entropy is bounded \textit{between 0 and 1}, where a bigger number means higher entropy of sensor data. 

The summary of our similarity metrics and entropy estimation is provided in~\autoref{tab:sim-entr-metr}.

\smallskip
\noindent
\textbf{End-to-end Comparison.}
To evaluate how \name can benefit existing solutions, we \textit{prototype} a state-of-the-art \gls{zip} \textit{and} \gls{zia} scheme, \textit{comparing} their error rates as well as completion time on the context data (1) impacted by \name and (2) affected by humans. 

\subsection{Results: Audio Context}
\label{sub:audio-res}
We present our findings on the similarity and entropy of audio context in the home and office scenarios. 

\begin{figure}
	\centering
	\begin{subfigure}[b]{.495\linewidth}
		\centering
		\includegraphics[width=\linewidth]{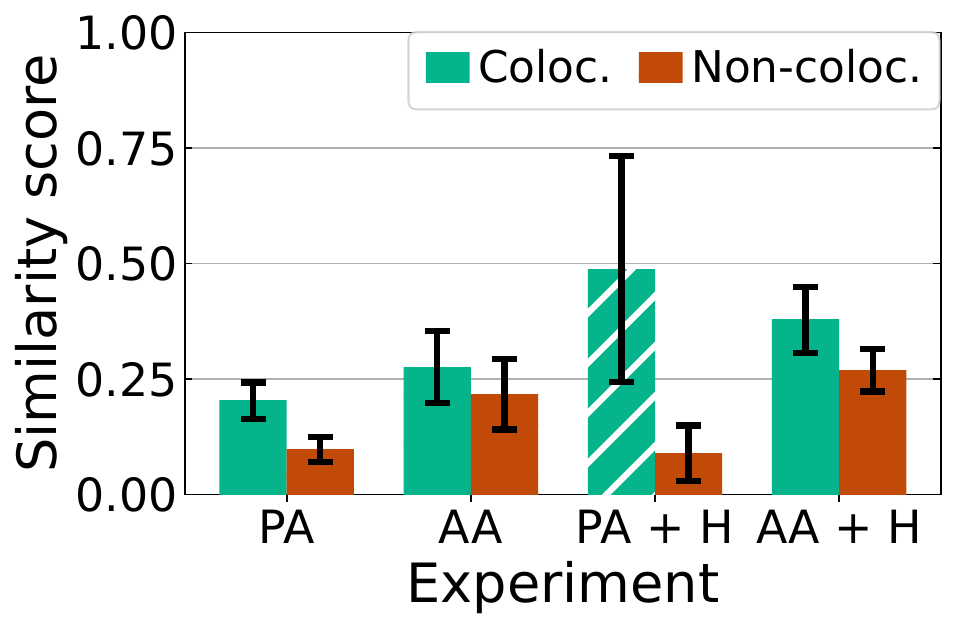} 
		\caption{Similarity scores (home)}
		\label{sf:distance-audio-Xcorr}
	\end{subfigure}
	\begin{subfigure}[b]{.495\linewidth}
		\centering
		\includegraphics[width=\linewidth]{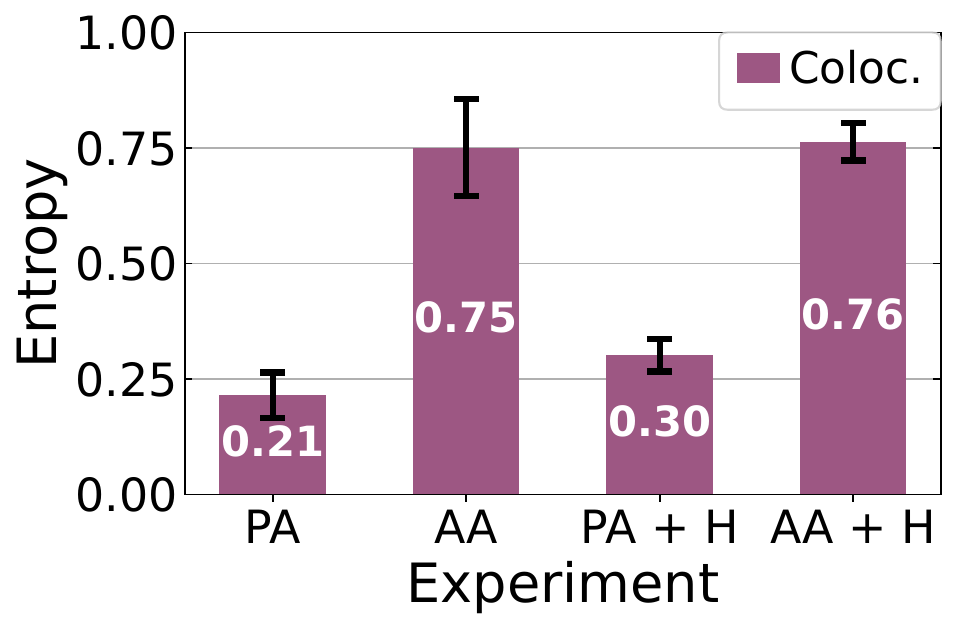}
		\caption{Audio entropy estimation (home)}
		\label{sf:entropy-audio-home}
	\end{subfigure}
	\caption{Evaluation of context (a) similarity and (b) entropy for audio data.}
	\vspace{-3.5ex}
\end{figure}

\smallskip
\noindent
\textbf{Similarity.} \autoref{sf:distance-audio-Xcorr} depicts the similarity scores (y-axis) of colocated and non-colocated devices in our four experiments (x-axis, cf.~\autoref{tab:exp-settings}) for the home scenario. 
We see two main trends here: (1) \textit{without} \name, the passive attack (\textit{PA}) is avoided by a narrow margin, while the active attack (\textit{AA}) \textit{succeeds};\footnote{The distributions of colocated and non-colocated similarity scores significantly overlap, as indicated by intersecting error bars of \textit{AA} in~\autoref{sf:distance-audio-Xcorr}.} (2) using \name allows preventing the passive attack (\textit{PA + H}) and mitigating the active attack (\textit{AA + H}). 
In the office scenario, the similarity results are alike. 

The similarity scores of~\autoref{sf:distance-audio-Xcorr} are obtained on the audio snippets of 60 seconds (cf.~\autoref{sub:eval-meth}).
With the shorter snippets of 10 and 30 seconds, we have the same trends, but the similarity of colocated devices is 15--30\% \textit{higher} while it \textit{grows by} up to 50\% for non-colocated devices during active attacks, i.e., \textit{AA} and \textit{AA + H}. 
Hence, longer audio recordings can help \gls{zip} and \gls{zia} schemes to resist attacks, corroborating findings from prior work~\cite{Fomichev:2019perils}. 

Exploring the high deviation of similarity scores between colocated devices when \name executes (cf. \textit{PA + H} in \autoref{sf:distance-audio-Xcorr}), 
we find that it stems from \textit{varying perception} of higher-frequency audio spectrum by obstructed devices. 
The similarity score captures audio spectrum till around 4.5 kHz. 
In the home scenario, the one device which is blocked by the bed, as presented in ~\autoref{fig:exp-setup},  senses slightly distorted audio, injected by the built-in speaker of a Roborock S5, than other colocated devices, causing discrepancies in similarity scores. 
We do not observe such behavior in the office scenario, however it reveals obstacles to using higher audio frequencies for audio injection in \name.

In~\autoref{sf:distance-audio-Xcorr}, we conduct the active attacks utilizing an affordable A2 speaker that plays \textit{constant} sine-waveform noise at 150 Hz (cf.~\autoref{sub:stim-alg}).
We use such low frequency, as it allows the noise to easily propagate into the environment of colocated devices without saturating the adversarial microphone situated near the A2 speaker. 
While this active attack already succeeds (cf. \textit{AA} in~\autoref{sf:distance-audio-Xcorr}), we further advance it by generating noise on various frequencies, i.e., \textit{sequentially} on each one-third octave band starting from 40 Hz to 20 kHz.
Such resulting noise resembles a frequency \textit{staircase}.  

\begin{figure}
	\centering
	\begin{subfigure}[b]{.495\linewidth}
		\centering
		\includegraphics[width=\linewidth]{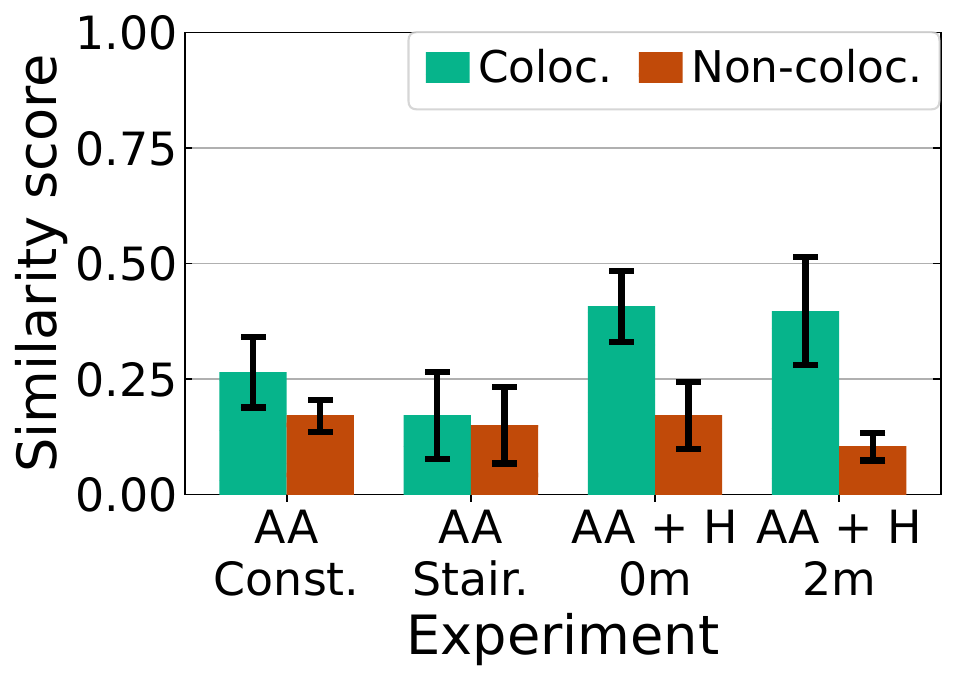} 
		\caption{Similarity scores for different adversarial signals and distances (office)}
		\label{sf:distance-audio-Xcorr-insight1}
	\end{subfigure}
	\begin{subfigure}[b]{.495\linewidth}
		\centering
		\includegraphics[width=\linewidth]{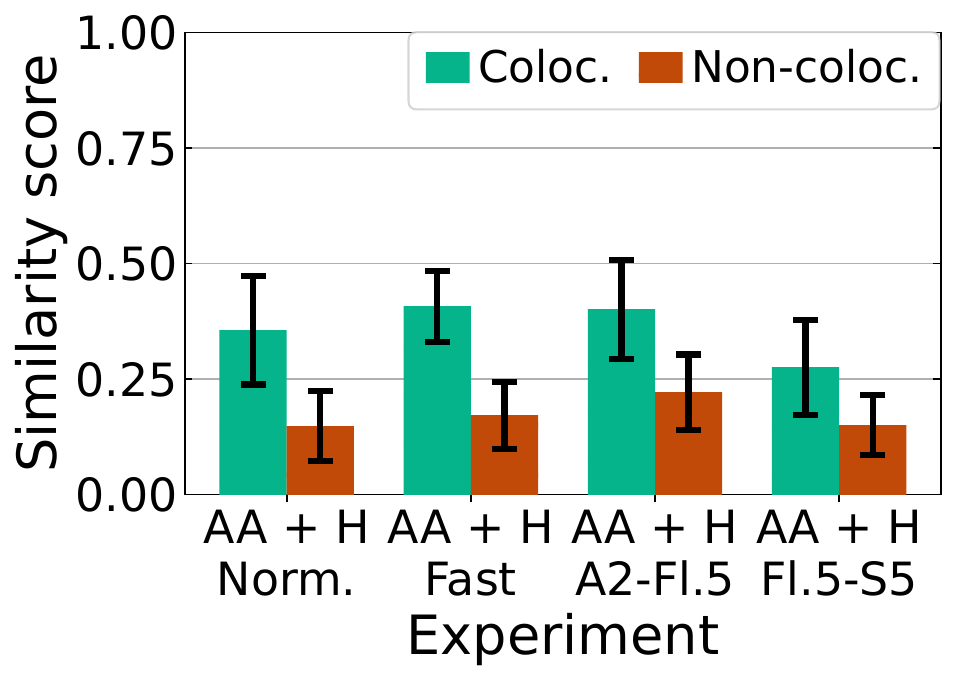}
		\caption{Similarity scores for different injection speed of \name and hardware (office)}
		\label{sf:distance-audio-Xcorr-insight2}
	\end{subfigure}
	\caption{Detailed analysis of audio context similarity under active attacks.}
	\vspace{-3.5ex}
\end{figure}

\autoref{sf:distance-audio-Xcorr-insight1} presents the similarity scores for the \textit{AA} and \textit{AA + H} cases in the office scenario. 
We see that the active attack using the staircase strategy is far more efficient than with the constant noise (cf. \textit{AA}: Const. vs. Stair. in~\autoref{sf:distance-audio-Xcorr-insight1}). 
Still, \name prevents the staircase attack which becomes \textit{impractical} if we increase the distance between the adversarial speaker and colocated devices by two meters, as depicted by \textit{AA + H}: 0m vs. 2m in~\autoref{sf:distance-audio-Xcorr-insight1}.

To assess the impact of \name's audio injection parameters on the efficacy of active attack prevention, we \textit{halve} the frequency of audio occurrence from 1--3 to 0.5--1.5 minutes (cf.~\autoref{tab:stim-params}).  
\autoref{sf:distance-audio-Xcorr-insight2} shows that such a change alone can not only increase the similarity scores of colocated devices but also make them more consistent (lower error bars), thwarting active attacks, i.e.,    \textit{AA + H}: Norm. vs. Fast. 
In the same fashion, we evaluate how \textit{high-quality speakers} affect the efficiency of both \name and the active attack. 
For this, we compare the similarity scores of colocated and non-colocated devices when (1) \name uses the better Flip 5 speakers while the adversary---A2; (2) vice versa: with the adversary utilizing Flip 5, whereas \name relies on the built-in Roborock S5 speaker. 
Indeed, higher-quality speakers benefit both \name and adversaries by either mitigating or aggravating the active attacks (cf. \textit{AA + H}: A2--Fl.5 vs. Fl.5--S5 in~\autoref{sf:distance-audio-Xcorr-insight2}).

\smallskip
\noindent
\textbf{Entropy.}
\autoref{sf:entropy-audio-home} provides the estimated audio entropy (y-axis) of colocated devices in our four experiments (x-axis)---listed in~\autoref{tab:exp-settings}---for the home scenario. 
We see that running \name \textit{boosts} the amount of entropy by over 40\%, i.e., from 0.21 in \textit{PA} to 0.30 in \textit{PA + H}. 
This allows \gls{zip} and \gls{zia} schemes to reduce their completion time by the \textit{same factor}. 
Interestingly, that during active attacks (cf. \textit{AA} and \textit{AA + H} in~\autoref{sf:entropy-audio-home}) performed using constant noise at 150 Hz, the amount of entropy is more than doubles, as compared to the \textit{PA + H} case. 
Studying this phenomenon, we find that the A2 speaker (employed by the adversary) vibrated when it played the noise while lying down on the tiled floor, producing additional clinking sounds. 
We attribute the high entropy figures of \textit{AA} and \textit{AA + H} cases to such sound artifacts. 

We obtain a more consistent picture of assessed audio entropy in the office scenario. 
Recall that the office's \textit{PA} experiment includes two persons, who actively influence the audio context (cf.~\autoref{sub:exp-setup}). 
Thus, our entropy estimates for the \textit{PA} and \textit{PA + H} cases account for 0.47 and 0.46, respectively. 
This result is crucial, indicating that \name can \textit{replace} intense human interaction in producing high-entropy context for \gls{zip} and \gls{zia}. 
In the office scenario, we do not notice any sound artifacts, seen in the home, resulting in consistent entropy figures of 0.48 and 0.54 in the active attack cases of \textit{AA} and \textit{AA+ H}, respectively.

Following our evaluation for the audio similarity, we find that utilizing high-quality speakers (i.e., Flip 5 vs. S5) allows \name to increase the entropy from the office's \textit{PA + H} estimate by an extra 20\%.
While injecting the audio stimulus twice as often, boosts the same entropy estimate by an added 40\%, already with the built-in Roborock S5 speaker. 

\subsection{Results: Illuminance and RGB Context}
\label{sub:light-res}
We report on the similarity and entropy of illuminance as well as RGB context found in the home and office scenarios.  

\begin{figure}
	\centering
	\begin{subfigure}[b]{.495\linewidth}
		\centering
		\includegraphics[width=\linewidth]{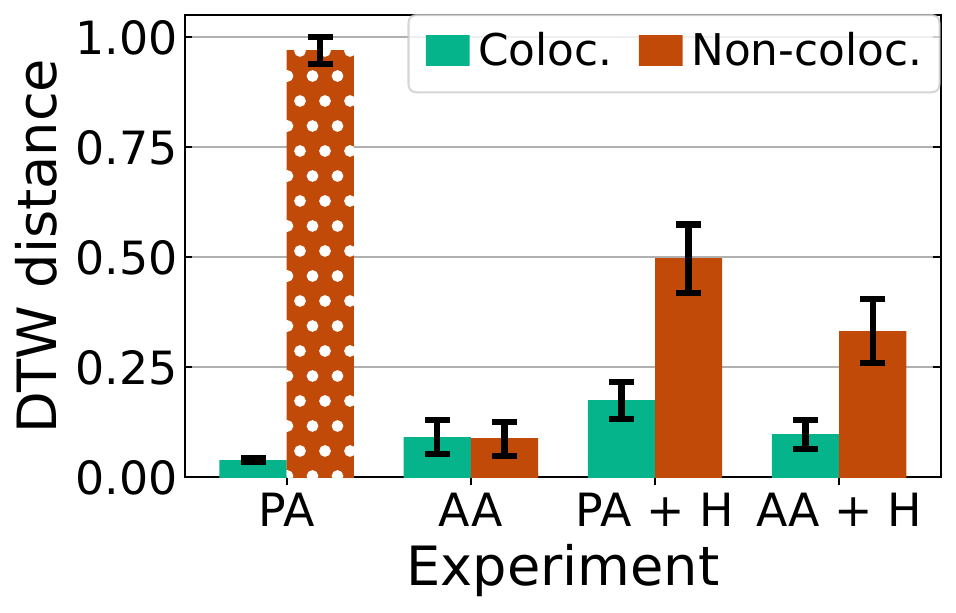} 
		\caption{RGB \gls{dtw} distances (home)}
		\label{sf:distance-light-rgb}
	\end{subfigure}
	\begin{subfigure}[b]{.495\linewidth}
		\centering
		\includegraphics[width=\linewidth]{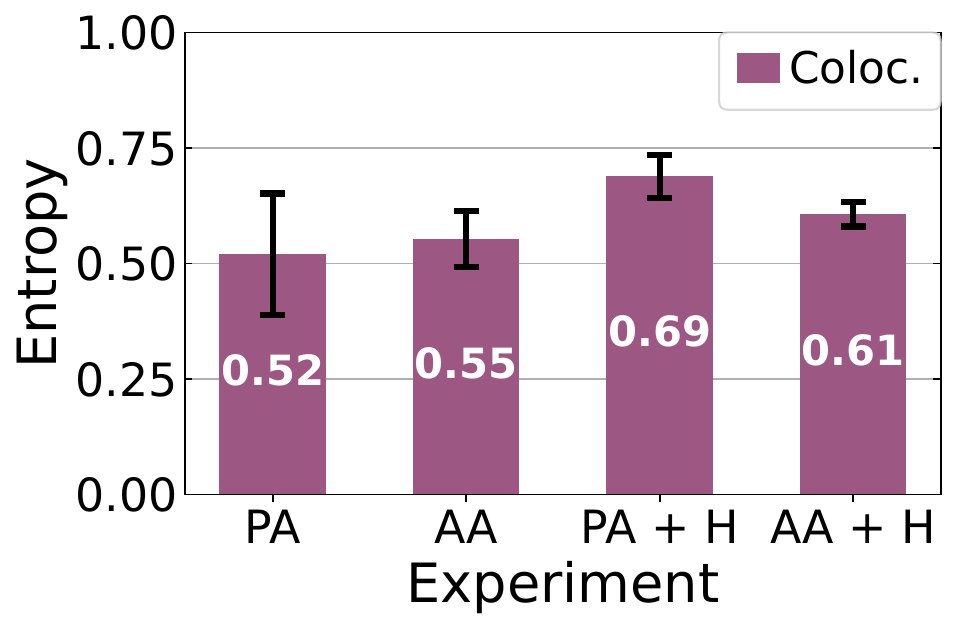}
		\caption{RGB entropy estimation (office)}
		\label{sf:entropy-light-rgb}
	\end{subfigure}
	\caption{Evaluation of context (a) similarity and (b) entropy for RGB data.}
	\vspace{-3.5ex}
\end{figure}

\smallskip
\noindent
\textbf{Similarity.}
\autoref{sf:distance-light-rgb} shows the \gls{dtw} distances (y-axis) between colocated and non-colocated devices for the RGB context in our four experiments (x-axis, cf.~\autoref{tab:exp-settings}) for the home scenario; the illuminance results are similar.  
We see that the passive attack \textit{fails} due to varying lighting conditions in- and outside the apartment room (e.g., daylight vs. electric), while the active attack, conducted with a colored lamp mounted on top of a half-open door, \textit{succeeds}---as depicted by \textit{PA} vs. \textit{AA} in~\autoref{sf:distance-light-rgb}.
\name prevents such an active attack (cf. \textit{AA + H} in~\autoref{sf:distance-light-rgb}), yet it slightly enlarges context dissimilarity (i.e., bigger \gls{dtw} distances) among colocated devices, suggesting that the coverage and directionality of light injections need to be considered. 

In the office scenario, we observe the same \gls{dtw} distance trends as in the home for both illuminance and RGB context, except that the active attack without \name, i.e., the \textit{AA} case, \textit{does not} succeed.
This happens because the adversarial light stimuli injected towards colocated devices are blocked by obstacles in the office, like displays. 

We study the \gls{dtw} distance on snippets of 15, 30, 60, 90, 120, and 300 seconds for illuminance and RGB data in both home and office scenarios. 
Our results show that the snippets of \textit{15--60 seconds} universally attain smallest \gls{dtw} distances, thus capturing the context similarity most efficiently. 
Hence, \gls{zip} and \gls{zia} schemes that rely on illuminance and RGB data can most benefit from these recording sizes, given the current light injection parameters of \name (cf.~\autoref{tab:stim-params}).  

\smallskip
\noindent
\textbf{Entropy.}
\autoref{sf:entropy-light-rgb} depicts the found RGB entropy (y-axis) of colocated devices in our four experiments (x-axis, cf.~\autoref{tab:exp-settings}) for the office scenario. 
We see that \name \textit{raises} the amount of entropy by more than 30\%---from 0.52 to 0.69 in the \textit{PA} and \textit{PA + H} cases, respectively; allowing \gls{zip} and \gls{zia} schemes using such data to \textit{proportionally} shorten their completion time. 
In the \textit{AA} and \textit{AA + H} cases, the estimated RGB entropy behaves as expected, according to presence or absence of context stimuli, whether from \name or the adversary. 
Our results for the illuminance entropy are alike. 

In the home scenario, the entropy estimates for RGB and illuminance follow the same trends as in the office while attaining comparable figures. 

\subsection{Results: \gls{co2} Context}
\label{sub:co2-res}
We provide the results for the \gls{co2} context similarity and entropy obtained in our home and office scenarios. 

\begin{figure}
	\centering
	\begin{subfigure}[b]{.495\linewidth}
		\centering
		\includegraphics[width=\linewidth]{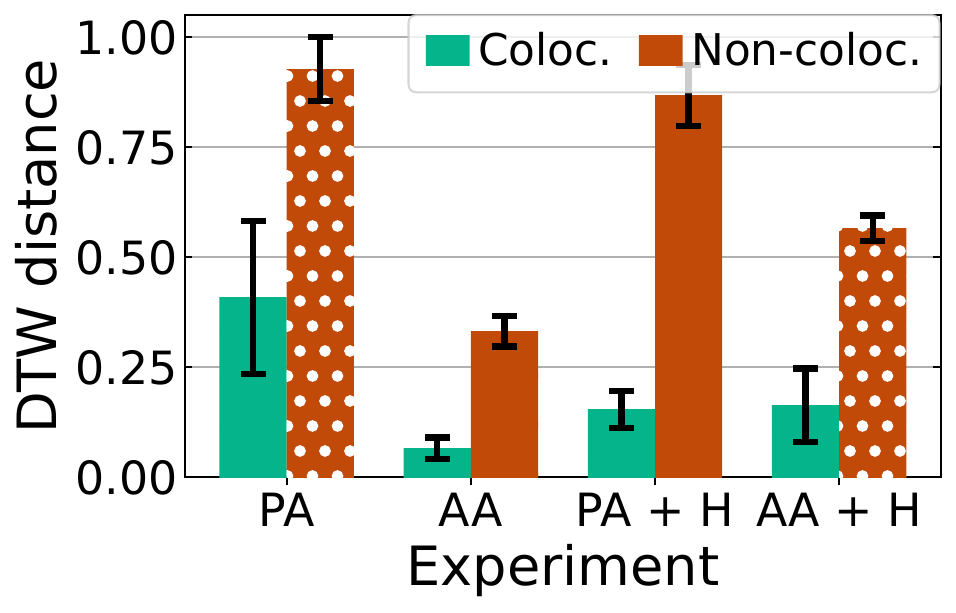} 
		\caption{\gls{co2} \gls{dtw} distances (office)}
		\label{sf:distance-gas-co2}
	\end{subfigure}
	\begin{subfigure}[b]{.495\linewidth}
		\centering
        \includegraphics[width=\linewidth]{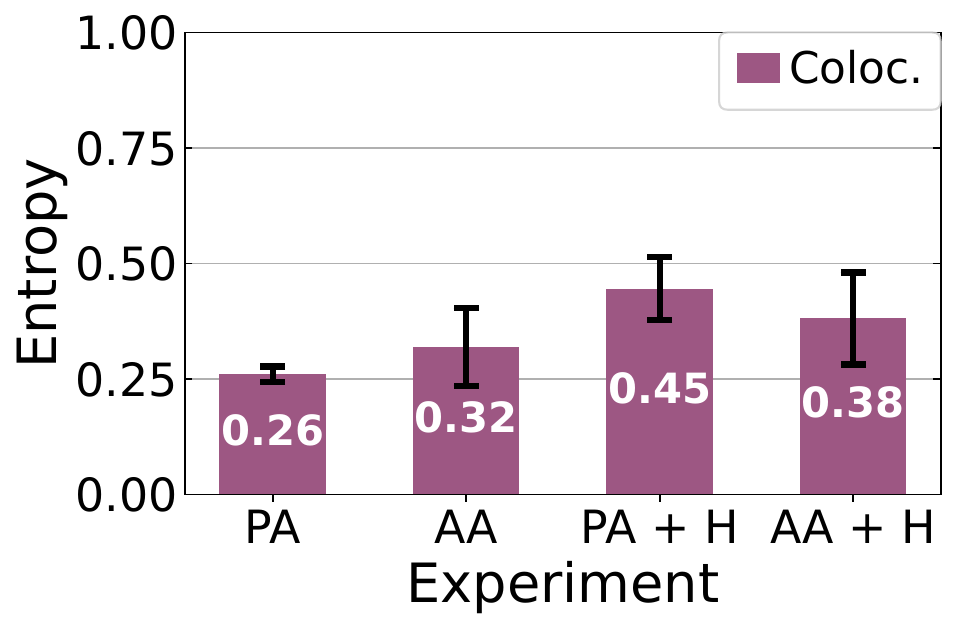}
		\caption{\gls{co2} entropy estimation (home)}
		\label{sf:entropy-gas-co2}
	\end{subfigure}
	\caption{Evaluation of context (a) similarity and (b) entropy for \gls{co2} data.}
	\vspace{-3.5ex}
\end{figure}

\smallskip
\noindent
\textbf{Similarity.}
\autoref{sf:distance-gas-co2} shows the \gls{dtw} distances (y-axis) between colocated and non-colocated devices for our four experiments (x-axis, cf.~\autoref{tab:exp-settings}) in the office scenario. 
We observe that the passive attack (\textit{PA}) \textit{fails}, as the ambient \gls{co2} levels in- and outside the office room differ significantly being affected by various factors, e.g., temperature and ventilation.
Such factors also contribute to \textit{reduced} \gls{co2} similarity between colocated devices, as indicated by the bigger \gls{dtw} distance and higher error bars (cf. \textit{PA} in \autoref{sf:distance-gas-co2}).
We see that the active attack (\textit{AA}), carried out by blowing air using a pedestal fan via a half-open door, is also \textit{unsuccessful}.
Still, non-colocated \gls{dtw} distances of \textit{AA} are much smaller than in the \textit{PA} case, suggesting the feasibility of active attacks on the \gls{co2} context, like through more advanced actuators, e.g., air pumps or industrial humidifiers. 

Using \name \textit{benefits} the \gls{co2} context by (1) making it \textit{more similar} and \textit{consistent} between colocated devices, as seen from smaller and less deviating \gls{dtw} distances of the \textit{PA + H} case than
\textit{PA} in~\autoref{sf:distance-gas-co2}; and (2) further \textit{raising the bar} for active attacks (cf. \textit{AA + H} vs. \textit{AA} in~\autoref{sf:distance-gas-co2}).

In the home scenario, the \gls{dtw} distance trends as well as figures are akin to the office results.

We explore the \gls{dtw} distance on snippets of 30, 60, 150, 300, 600, and 900 seconds for \gls{co2} data 
in both home and office scenarios. 
Our findings suggest that the snippets in the range \textit{from 300 to 600 seconds} universally result in smallest \gls{dtw} distances, thus capturing \gls{co2} context most similarly. 
Hence, such recording sizes can significantly profit \gls{zip} and \gls{zia} schemes utilizing \gls{co2} data, for the existing parameters of \gls{co2} injection in \name (cf.~\autoref{tab:stim-params}).
 
\smallskip
\noindent
\textbf{Entropy.} 
\autoref{sf:entropy-gas-co2} gives the assessed \gls{co2} entropy (y-axis) of colocated devices for our four experiments (x-axis, cf.~\autoref{tab:exp-settings}) in the home scenario. 
\name attains the \textit{entropy increase} of over 70\%, i.e., from 0.26 in \textit{PA} to 0.45 in \textit{PA + H}. 
This enables \gls{zip} and \gls{zia} schemes based on \gls{co2} data to \textit{comparably} decrease their completion time. 
We also see that the active attack, which does not succeed in terms of context similarity (cf.
\textit{AA + H} in~\autoref{sf:distance-gas-co2}), can slightly reduce the entropy boost provided by \name, as shown by the \textit{AA + H} case of~\autoref{sf:entropy-gas-co2}. 

In the office scenario, we observe similar entropy behavior with marginally higher figures, when using the same affordable Maxcio humidifier (cf.~\autoref{tab:exp-hardware}). 
With the more advanced iTvanila Mist, we can further raise the amount of entropy by an additional 20\% from the home's \textit{PA + H} estimate (cf.~\autoref{sf:entropy-gas-co2}), accounting for 0.54. 

\subsection{End-to-end Comparison with Prior Work}
\label{sub:eval-cmp}
We now \textit{compare how} a state-of-the-art \gls{zip}~\cite{Miettinen:2014} and \gls{zia} \cite{Karapanos:2015} scheme perform in terms of security, usability, and completion time on context data collected in two office scenarios: (1) occupied by humans~\cite{Fomichev:2019perils} and (2) devoid of them, but with \name working. 
To provide a meaningful comparison, we pick the schemes utilizing different context data, i.e., illuminance and audio.
In~\cite{Fomichev:2019perils}, these data are captured in three office rooms, encompassing four colocated devices as well as between one to three humans each, throughout eight hours. 
While our data are collected in one office with four colocated devices inside and no people for six hours (cf.~\autoref{sf:setup-office}). 

\smallskip
\noindent
\textbf{\gls{zip} Scheme.} We replicate \gls{zip} by \textit{Miettinen et al.}~\cite{Miettinen:2014} that relies on illuminance, using the scheme's implementation of \cite{Fomichev:2019perils}, allowing for a direct comparison. 
As reported by~\cite{Fomichev:2019perils}, this scheme achieves lowest error rates 
when one fingerprint bit (i.e., 0 or 1) is obtained from 30 seconds of illuminance data---we follow such parametrization in our comparison. 
To assign a 0- or 1-bit, the \gls{zip} scheme utilizes two thresholds, which are set to 10 and 0.1 by prior works~\cite{Miettinen:2014, Fomichev:2019perils}.  
We find these threshold values to be conservative, as they only allow capturing \textit{profound illuminance changes}, like turning on / off electric lighting or shading daylight.   
Hence, the fingerprints produced by this scheme contain mostly 0-bits, rendering it insecure; this limitation is also noted by~\cite{Fomichev:2019perils}.
We thus adjust the thresholds to 8 and 0.01, respectively, making it possible to catch less extreme changes in illuminance (e.g., brightness alteration) within our fingerprints. 

\begin{table}
\scriptsize
\centering
	\caption{Evaluation of \gls{zip} scheme from~\cite{Miettinen:2014} on illuminance data affected by (1) humans~\cite{Fomichev:2019perils} and (2) \name  in the office scenario.}
	\label{tab:cmp-zip}
	\begin{tabular}{llccc}
		\toprule
		\multirow{2}[2]{*}{\shortstack[l]{Illum. \\ data}} & \multirow{2}[2]{*}{\shortstack[l]{EER}}  &  \multicolumn{3}{c}{Fingerprint quality (colocated)} \\ 
		\cmidrule(rl){3-5}
		 &  & \% of 1-bits & Min-entropy & \% of \textit{balanced} keys\\
		 \midrule
		 \cite{Fomichev:2019perils} &  0.54  & 8.9 \textpm\ 0.6 & 0.004  & 3.2 \\
		 \name &  0.07  & 25.3 \textpm\ 4.1 & 0.26  & 9.1 \\
		 \bottomrule
	\end{tabular}
	\smallskip\centering
  	\center{EER -- Equal Error Rate. \\ Min-entropy is assessed in one bit; the balanced keys are 20 bits in size.} 
  	\vspace{-3.5ex}
\end{table}

\autoref{tab:cmp-zip} shows our evaluation results of \gls{zip} by Miettinen et al.~\cite{Miettinen:2014} obtained on illuminance data impacted by humans \cite{Fomichev:2019perils} and \name. 
We find that the \textit{average difference} in fingerprint similarity of colocated and non-colocated devices is 12.4\% for the \name data, while it reaches only 0.1\% on the illuminance from~\cite{Fomichev:2019perils}.
To assess the \textit{security and usability} of this scheme, we compute \gls{eer}, that is the intersection point of \gls{far} and \gls{frr}. 
The \gls{eer} thus balances security (\gls{far}) and usability (\gls{frr}) metrics, which are equally important for \gls{zip} and \gls{zia} schemes~\cite{xu2021key, conti2020context}.
We obtain \glspl{eer} of 0.07 and 0.54 on the data from \name and~\cite{Fomichev:2019perils}, respectively. 
Hence, \name \textit{greatly} improves the security and usability of the \gls{zip} scheme under consideration. 

We further explore the quality of fingerprints produced by this \gls{zip} scheme to evaluate not only its security---as \textit{unpredictability} of fingerprints---but also the pairing time. 
\autoref{tab:cmp-zip} reveals that fingerprints derived from the illuminance data of \cite{Fomichev:2019perils} contain less than 10\% of 1-bits, explaining the \textit{striking similarity} of colocated and non-colocated fingerprints, since both are mainly comprised of 0-bits. 
Even worse, these fingerprints are prone to include long sequences of 0-bits which are rarely interleaved with short series of consecutive 1-bits.  
Such fingerprints are easily predictable---we confirm this by assessing their \textit{min-entropy} in one bit using the NIST SP800-90B test suite~\cite{turan2018recommendation} (cf.~\autoref{tab:cmp-zip}). 
Recall that the low entropy of fingerprints leads to a prolonged pairing time~\cite{Fomichev:2021}. 

Inspecting fingerprints obtained on the \name data, we see that (1) the ratio of 1-bits exceeds 25\% while (2) there exist few long sequences of consecutive 0- and 1-bits. 
Thus, our fingerprints have \textit{65-times more entropy} in one bit, estimated with the NIST suite, than those derived from the illuminance data of~\cite{Fomichev:2019perils} (cf.~\autoref{tab:cmp-zip}). 
Hence, the pairing times, that are achievable on the data by \name and \cite{Fomichev:2019perils}, can  potentially differ by the same factor.   

The NIST entropy tests can be \textit{inaccurate} requiring a significant amount of data for a fair assessment~\cite{turan2018recommendation}. 
Hence, we also estimate the amount of \textit{balanced} keys, i.e., fingerprints of a fixed size containing roughly the same number of 0- and 1-bits. 
We set the size of such fingerprints to 20-bits, which suffices to pair securely, using state-of-the-art cryptographic protocols~\cite{Fomichev:2021}.
Moreover, we demand that balanced keys lie within the 40:60 ratio between 0- and 1-bits (or vice versa), following prior work~\cite{Bruesch:2019}. 
To obtain such balanced keys, we traverse our fingerprints with an overlapping sliding window, using the step of 4-bits. 
We see that the \name data allows producing \textit{three times more} balanced keys that the illuminance from~\cite{Fomichev:2019perils} (cf.~\autoref{tab:cmp-zip}), shortening the pairing time of the studied \gls{zip} scheme by the similar degree. 

\smallskip
\noindent
\textbf{\gls{zia} Scheme.} For \gls{zia} by \textit{Karapanos et al.} \cite{Karapanos:2015}, we already adapt the scheme implementation from~\cite{Fomichev:2019perils} (cf. Section~\ref{sub:sim-ent}), enabling a fair comparison.
Both our findings in~\autoref{sub:audio-res} and those of~\cite{Fomichev:2019perils} indicate that this scheme can best separate colocated and non-colocated devices on audio snippets of 60 seconds. 
Thus, we use the similarity scores obtained on such snippets to compute \glspl{eer} for audio data affected by humans \cite{Fomichev:2019perils} and \name. 
\autoref{tab:cmp-zia} shows that on the \name data, which is taken under active attacks, we achieve a lower \gls{eer} than for the audio from~\cite{Fomichev:2019perils}, where only passive attacks are considered (cf.~\autoref{sec:mod}). 
Thus, \name can \textit{further harden} \gls{zia} by Karapanos et al., improving its security and usability, even during active attacks.  

\begin{table}
\scriptsize
\centering
	\caption{Evaluation of \gls{zia} scheme from~\cite{Karapanos:2015} on audio data affected by (1) humans~\cite{Fomichev:2019perils} and (2) \name in the office scenario.}
	\label{tab:cmp-zia}
	\begin{tabular}{llc}
		\toprule
		\makecell[l]{Audio \\ data} & \makecell[l]{EER} & \makecell{\% of enough-power snippets \\ (colocated)} \\
		 \midrule
		 \cite{Fomichev:2019perils} &  0.08 & 90.0 \\
		 \name &  0.06 & 99.5 \\
		 \bottomrule
	\end{tabular}
		\smallskip\centering
  	\center{EER -- Equal Error Rate.} 
	\vspace{-3.5ex}
\end{table}

To discard audio snippets that have low entropy, thwarting predicable context attacks (e.g., in a quiet room), this scheme uses a \textit{power threshold} of 40 dB~\cite{Karapanos:2015, Fomichev:2019perils}.
Thus, the snippets, whose power is \textit{below} this threshold, are rejected prolonging the authentication time. 
To estimate this authentication time for \name data and audio from~\cite{Fomichev:2019perils}, we find how many audio snippets (60-second) of colocated devices have \textit{enough power}, lying above the power threshold. 
\autoref{tab:cmp-zia} reports that using such estimate, \name achieves almost \textit{10\% faster} authentication time for the studied \gls{zia} scheme.


\section{Discussion}
\label{sec:disc}
We present relevant discussion points for \name.

\smallskip
\noindent
\textbf{Generalizability.}
We design \name to generalize for various types of \gls{iot} actuators. 
The stimuli parameters of~\autoref{tab:stim-params}: frequency, duration, intensity, and pattern are \textit{generic}, hence they apply to different actuators, yet this parameter list is non-exhaustive. 
We also identify \textit{actuator-specific} parameters, e.g., color and spectrum, to consider the heterogeneity of \gls{iot} actuators.
Thus, \name can be \textit{extended} to other actuators, using a mixture of generic and specific stimuli parameters found in this work and by future research. 

To set the stimuli parameters of \name injection (cf. \autoref{tab:stim-params}), we \textit{need to} balance the capabilities of \gls{iot} actuators, i.e., how fast they can act, and the dynamics with which their stimuli affect context. 
Increasing the stimulus frequency, as shown in~\autoref{sub:audio-res}, boosts both the context similarity and entropy. 
However, one should consider that in the \textit{real-world} settings, \name must not disrupt the main routine of an actuator or consume too much power, depleting its battery. 

\smallskip
\noindent
\textbf{Deployment Considerations.}
We intend \name to operate in \textit{unattended} scenarios which are prevalent in the \gls{iot}, like a smart home without inhabitants during working hours. 
Still, in many scenarios humans are present, yet hardly affect the context, e.g., by sitting quietly or sleeping. 
In such cases, \name can help to produce high-entropy context, but it may be \textit{perceived as invasive} by people; we elaborate on this issue in the next discussion point. 
We envision \name can be tailored to work in human presence by using \textit{less obtrusive} context stimuli, e.g., ultrasound, and applying them in a \textit{directional manner}, like emitting light towards colocated devices. 
Such directionality can be achieved by utilizing mobile \gls{iot} actuators, e.g., robotic vacuum cleaners. 
Still, more research is required to study the feasibility of these methods. 

Another concern is the security of \name due to its \textit{reliance on} the \gls{prng}. 
\glspl{prng} are plagued with issues, producing numbers in a deterministic fashion~\cite{kietzmann2021guideline}.
Recent work accentuates this problem for \gls{iot} devices whose \glspl{prng} are \textit{commonly} insecure~\cite{tillmanns2020firmware}. 
Thus, one should consider such an issue when prototyping \name on real hardware.

Our findings indicate that the \textit{capabilities} of \gls{iot} actuators (e.g., speaker quality) and their \textit{placement} (e.g., light coverage) play an important role for the efficiency of \name. 
Hence, to leverage \name, the actuators should be deployed such that their stimuli could reach colocated devices, performing \gls{zip} or \gls{zia}, as seamlessly as possible. 

\smallskip
\noindent
\textbf{Insight: \name and Human Presence.}
Our extra experiments, where we have \textit{two persons} within the office scenario whilst \name \textit{works} (cf.~\autoref{sub:exp-setup}), inform this discussion point. 
First, we check if human presence may interfere with \name, reducing context similarity and / or entropy.
We see \textit{no} significant difference between the results of these experiments and those provided in Sections  \ref{sub:audio-res}--\ref{sub:co2-res}, for all types of context data that we study. 
Thus, \name \textit{maintains its efficiency} in the face of human presence. 

Second, we ask our participants, who are two males of 32 and 28 years old as well as tech-savvy, about their perception of \name in terms of intrusiveness. 
Interestingly, both respondents found \name to be \textit{less intrusive} than we had anticipated. 
Specifically, they \textit{hardly noticed} the humidifier (blowing vapor) that worked ``soundlessly'' according to one person, while another---had a similar humidifier at home and was used to it. 
Our participants reported the same experience with smart lights, clarifying that the blinking bulbs \textit{did not bother} them, as they were not the only light source in the room, since the main ceiling lights were on. 

Unsurprisingly, both respondents rated the audio injection as the \textit{most intrusive}. 
When inquired about the level of intrusiveness, one person compared it to a busy day in the office, where visitors drop by, while another---imagined a gathering of several people nearby.  
We further asked for how long our participants are able to endure the audio injection, providing they fully understand the \name purpose. 
Both agreed that \textit{2--4 minutes} of audio injection are acceptable.

We consider the above experience of our participants with \name to be the first step to understanding of how users view it, paving the way for the \name deployment.  

\smallskip
\noindent
\textbf{Limitations and Future Work.}
In our audio injection based on speech, we do not consider \textit{vocal registers} occupying different frequencies (e.g., fry--low vs. whistle--high). 
This can diversify speech produced by \name, \textit{raising entropy} of the audio context. 
Now, we generate random speech utilizing the state-of-the-art \gls{rnn} (cf.~\autoref{sub:stim-alg}).
Yet, in this \gls{rnn}, we only use English language, hence adding more languages and mixing them would make the produced speech more \textit{unpredictable}. 
As an alternative to \glspl{rnn}, we can leverage the GPT-4 mechanism\footnote{\url{https://openai.com/product/gpt-4}} for generating random text---from which \name will synthesize the speech. 

A different approach to audio injection can rely on \textit{white noise}, where the phase of different frequencies is controlled by the \gls{prng}, to attain higher entropy of audio context.
We leave the investigation of such an approach to future work. 

In our evaluation, we were unable to perform a successful active attack on the \gls{co2} context while showing its feasibility (cf. \autoref{sub:co2-res}). 
We envision this limitation to be addressed, as part of future research, by leveraging more \textit{advanced actuators}, like air pumps and industrial humidifiers.  

Another avenue for improvement is to develop a \textit{generic method} to select bins in our entropy estimation for different types of context data (cf.~\autoref{sub:sim-ent}). 
So far, we rely on data properties, like range and behavior, to choose the number of bins. 
This allows us to evaluate the relative change in entropy per sensor data type (e.g., audio), but \textit{does not} compare the entropy of various data, like \gls{co2} vs. illuminance. 
Such bin selection is prone to \textit{over- or underestimate} the entropy if the bins are chosen without considering certain artifacts of data behavior, as seen for audio in~\autoref{sub:audio-res}. 
Note that devising the generic method to select bins can be inspired by~\cite{ross2014mutual, west2021moonshine}. 


\glsresetall

\section{Conclusion}
\label{sec:concl}
\Gls{zip} as well as \gls{zia} allow \gls{iot} devices to establish and maintain secure communication \textit{without user assistance}, leveraging their \textit{ambient context}, like audio. 
The amount of entropy in this context is \textit{decisive} for the security and completion time of \gls{zip} and \gls{zia} schemes. 
Such context entropy is often low, especially in \textit{unattended scenarios} (e.g., empty smart home) that are prevalent in the \gls{iot}, threatening both security and utility of the schemes. 
To address these issues, we propose \name, a novel approach that utilizes
off-the-shelf \gls{iot} \textit{actuators} and their \glspl{prng} for producing high-entropy context. 
We implement \name on commodity actuators, i.e., smart speakers, lights, humidifiers, and conduct real-world experiments, showing the capability of \name to prevent advanced \textit{active attacks} on \gls{zip} and \gls{zia} schemes, while \textit{speeding them up} by up to two times. 


\section{Acknowledgments}
\label{sec:ack}
We would like to thank the anonymous reviewers for their insightful comments. 
This work has been co-funded by the Research Council of Norway as part of the project Parrot (311197) and the Deutsche Forschungsgemeinschaft (DFG) within the Collective Resilient Unattended Smart Things (CRUST) project as part of the DFG Priority Program SPP 2378 -- Resilient Worlds.

%
%

%
%
\balance
\bibliographystyle{abbrv}
\bibliography{bibliography}  
\end{document}